\documentclass[useAMS,usenatbib,usegraphicx]{mn2e}

\title[Hot subluminous stars in {\it GALEX}]{A selection of hot subluminous stars in the {\it GALEX} survey\\
I. Correlation with the Guide Star Catalog
\thanks{Based on observations made with ESO telescopes at the La Silla Paranal Observatory
under programmes 82.D-0750, 83.D-0540, and 085.D-0866.}}

\author[S. Vennes et al.]{S. Vennes$^{1}$\thanks{E-mail: vennes@sunstel.asu.cas.cz (SV); kawka@sunstel.asu. cas.cz 
(AK); nemeth@sunstel.asu.cas.cz (PN)} \thanks{
Visiting Astronomer, Kitt Peak National Observatory, National Optical Astronomy Observatory,
which is operated by the Association of Universities for Research in Astronomy (AURA)
under cooperative agreement with the National Science Foundation.},
A. Kawka$^{1}$\footnotemark[2]\footnotemark[3], and P. N\'emeth$^{1,2}$\footnotemark[2]\\
$^{1}$Astronomick\'y \'ustav AV \v{C}R, Fri\v{c}ova 298,CZ-251 65 Ond\v{r}ejov,
Czech Republic\\
$^{2}$Department of Physics and Space Sciences, 150 W. University Blvd, Florida Institute of Technology, Melbourne, FL 32901, USA
}

\begin{document}
\date{}

\pagerange{\pageref{firstpage}--\pageref{lastpage}} \pubyear{2010}

\maketitle

\label{firstpage}

\begin{abstract}
We assembled a catalogue of bright, hot subdwarf and white dwarf stars extracted from 
a joint ultraviolet, optical, and infrared source list.
The selection is secured using colour criteria that correlate 
well with effective temperatures $T_{\rm eff}\ga 12,000$ K.
We built a $N_{\rm UV}-V$ versus $V-J$ diagram for $\ga 60,000$ bright sources using the
{\it Galaxy Evolution Explorer} ({\it GALEX}) $N_{\rm UV}$ magnitude ($N_{\rm UV}<14$), and the associated 
Guide Star Catalog (GSC2.3.2) photographic quick-V magnitude and the
Two Micron All Sky Survey (2MASS) J and H magnitudes. This distillation process delivered a
catalogue of $\approx 700$ sources with $N_{\rm UV}-V<0.5$ comprising $\sim160$ known
hot subdwarf stars and another $\sim 60$ known white dwarf stars. A reduced proper-motion diagram
built using the proper-motion measurements extracted from the Naval Observatory 
Merged Astrometric Dataset allowed us to identify an additional $\sim 120$ new hot subdwarf candidates
and $\sim 10$ hot white dwarf candidates. We present a spectroscopic study of a subset of 52 subdwarfs, 48 of them
analysed here for the first time, and with nine objects brighter than $V\sim12$. 
Our sample of spectroscopically confirmed hot subdwarfs comprises ten sdO stars and 42 sdB stars suitable 
for pulsation and binary studies. 
We also present a study of 50 known white dwarfs selected in the {\it GALEX} survey and six new white dwarfs from our catalogue of 
subluminous candidates. 
Ultraviolet, optical, and infrared synthetic magnitudes employed in the selection and analysis of white dwarf stars
are listed in appendix.
\end{abstract}


\section{Introduction}

Surveys of blue- or ultraviolet (UV)-excess objects are a rich source of hot subluminous stars and
are beneficial to the study of hot subdwarf and white dwarf stars. 
Following the pioneering work of \citet{hum1947}, colourimetric surveys, such as the 
northern and southern Tonantzintla survey \citep[TON/TON-S, see][]{iri1957} and
the Palomar-Haro-Luyten survey \citep{har1962} 
provided for many years the source lists for spectroscopic observations and analysis of
faint blue stars at high-Galactic latitude \citep[e.g.,][]{gre1966,gre1974}.

The spectroscopic observation and spectral classification of samples of subluminous stars
\citep[e.g.,][]{hum1947,fei1958,cow1958,sle1961,ber1963}
helped \citet{gre1974} define the extended (or extreme) horizontal branch (EHB) in the HR diagram.
Their diagram comprising 189 faint blue stars shows a
sequence of hot white dwarfs ranging from $T_{\rm eff}\approx14,000$ to 56,000 K and overlapping with
a sequence of sdO stars in the higher range of temperatures and luminosities ($\log{L/L_\odot}\ga 0$).
The evolutionary tracks of \citet{pac1971a,pac1971b} 
suggested for the first time that helium stars with masses in the range $0.5-0.85$ $M_\odot$ and without hydrogen envelopes 
evolve from the EHB toward the white dwarf cooling sequence. However, based on EHB space density
and birthrate estimates, \citet{heb1986} concluded that 
the EHB channel contributes only a small percentage of objects on the
white dwarf cooling sequence with the majority following the channel connecting planetary nebula nuclei
to the white dwarfs \citep{dri1985}.

\citet{gre1974} also found that most EHB stars have weak helium and weak heavy element lines, hence the label population II or ``halo", 
and are classified as sdB stars in contrast with the helium-rich sdO stars. In addition, the 
luminosity-to-mass ratio $L/M$, or its reciprocal $g\theta^4$ (where $\theta=5040\,K/T_{\rm eff}$), 
appeared approximately constant among these objects. Applying a model atmosphere
analysis to their sample, Greenstein \& Sargent extracted temperature ($T_{\rm eff}$) and surface 
gravity ($\log{g}$) measurements and, since $L/M\propto T_{\rm eff}^4/g$, they estimated 
$L/M\approx 68\ L_\odot/M_\odot$, or $L=46\,L_\odot$ for a representative mass of $\approx0.66\,M_\odot$. 
The amount of hydrogen present in the star was unknown although they surmised that
the nearly constant luminosity-to-mass ratio implied a thin hydrogen envelope.

The ground-based blue-excess surveys were soon followed by space-borne UV surveys.
The TD-1 satellite conducted a systematic, all-sky survey \citep{tho1978,car1983} that included
the Galactic plane. The survey was conducted in four bands, one photometric band centred on 2700 \AA\ and
three low-resolution spectroscopic bands centred on 1565, 1965, and 2365 \AA. 
The original catalogue of \citet{tho1978} includes 31215 entries with the catalogue prefix ``TD1'', out of which \citet{car1983} 
listed 464 objects with UV colours of spectral types B4V or earlier. Among these objects, 47 were labelled as subdwarfs (``sd''). 
The spectroscopic identifications were secured optically. For example, \citet{ber1980} listed 28 hot subluminous stars, 
thirteen of them new objects listed with the ``UVO'' catalogue prefix.

The sdB and sdO stars detected in the TD-1  have visual magnitudes ranging from 8 to 12 and are, to this day,
the brightest stars of their class. For example, the brightest star in the sample is the sdO HD~49798 \citep{jas1963,mer2009} 
with a UV flux at 2740 \AA\ of $f_\lambda =1.6\times10^{-11}$ erg\,cm$^{-2}$\,s$^{-1}$\,\AA$^{-1}$,
or $m_{\rm AB}(2300{\rm \AA})=7.8$ mag in the AB magnitude system\footnote{$m_{\rm AB} = -2.5\log{f_\nu}-48.6$.}. The next two brightest evolved stars
are HD~76431 \citep{ber1980,ram2001} and BD+75~325 \citep{gou1957} with $m_{\rm AB}(2300{\rm \AA})=8.9$ mag.
Few hot white dwarfs like UVO2309+10 \citep[$=$BPM~97895, GD~246;][]{luy1963,gic1965} were originally listed by \citet{car1983}, but TD-1 UV photometry 
eventually led to discoveries such as the
white dwarf companions to the F7~II star HR~3643 and the F4~V star 56~Persei \citep{lan1996}.
In summary, the TD-1 UV catalogue includes objects with $m_{\rm AB}(2300{\rm \AA})\le 11.5$ mag,
and, after early-type main-sequence stars, it is populated mostly with hot sdB and sdO stars.

Furthermore, the Orbiting Astronomical Observatory-2
\citep[OAO-2,][]{cod1980} delivered a catalogue of 531 pointed UV sources that contained few white dwarfs \citep[e.g., Feige24, ][]{hol1976}
and $\approx 20$ hot subdwarfs.
The Astronomical Netherlands Satellite (ANS) also produced a catalogue of 3573 pointed UV
sources \citep{wes1982} listing $\approx 10$ hot white dwarfs, analysed by \citet{wes1978}, and about three times as many hot subdwarfs. 
These experiments pre-selected bright subdwarfs with $f_\lambda \ga 10^{-13}$ and $4\times10^{-14}$ erg\,cm$^{-2}$\,s$^{-1}$\,\AA$^{-1}$ 
corresponding to $m_{\rm AB}(2300{\rm \AA})\la13.4$ and $\la$14.4, respectively.
Although the OAO-2 and ANS reached fainter magnitudes, the all-sky coverage of the TD-1 survey allowed for
the systematic identification of new subluminous stars.
Ultraviolet flux measurements of
subluminous stars are also listed in the Midcourse Space Experiment (MSX) UV catalogue \citep{new2004}.

Large catalogues of subluminous stars were first built based on extensive blue and UV-excess
surveys such as the Palomar-Green (PG) survey \citep{gre1986}, the Kitt-Peak-Downes (KPD) survey \citep{dow1986},
the Edinburgh-Cape (EC) survey \citep{kil1997},
the Montreal-Cambridge-Tololo (MCT) survey \citep{lam2000}, or the First Byurakan Survey \citep[FBS,][]{mic2008}.
These and similar catalogues provided the basis for our current understanding of EHB and post-EHB stars, as well as post-asymptotic
giant branch (AGB) stars.

Detailed spectroscopic studies based on the PG survey \citep{moe1990a,saf1994} 
placed hot sdB stars in the $T_{\rm eff}-\log{g}$ diagram close to the
evolution tracks of \citet{cal1989} for core helium-burning stars with very
thin hydrogen envelopes and masses near $0.5\,M_\odot$.
Recent spectroscopic investigations of the Hamburg-Schmidt (HS) subdwarf catalogue \citep{ede2003} 
using the evolutionary models of \citet{dor1993} confirmed the basic picture.
High mass-loss rates are possibly responsible for thinning 
the hydrogen envelope and preventing the star from climbing the asymptotic giant branch \citep{dcr1996}.
Other formation scenarios examined by \citet{han2002,han2003} involve a common-envelope
phase or an episode of Roche lobe overflow that may help remove the hydrogen envelope and direct the star toward the EHB as
originally proposed by \citet{men1976}, or involve the merger of two helium white dwarfs. 
Indeed, \citet{max2001}, \citet{mor2003}, and \citet{ede2005} found that a significant fraction of these stars reside in 
spectroscopic binary systems in
support of the scenario for the formation of sdB stars through close binary evolution.

The evolutionary paths leading to the formation of helium rich subdwarfs (sdO) are less certain although
the helium double-degenerate mergers \citep{han2002,han2003} or helium flash/mixing \citep{swe1974,swe1997a,swe1997b} 
may account for the paucity of hydrogen.
Analysis of helium-rich subdwarfs (sdO) from the PG survey \citep{dre1990,the1994} suggest that sdO stars 
evolved away from the EHB. Recently, \cite{str2007} draws further distinction between helium-deficient sdO stars
and helium-enriched sdO stars, with the former linked to sdB stars and the latter possibly formed in the merger
of helium white dwarfs, or via helium mixing.
Detailed abundance studies \citep[e.g.,][]{lan2004,ahm2007} offers insights into the formation of helium-rich subdwarf stars involving
helium mixing.
Based on a population study \citet{zha2009} conclude that many sdO stars evolve directly from sdB stars, 
and that only the more massive sdO stars ($M>0.5\,M_\odot$) are the outcome of white dwarf mergers.

The majority of white dwarfs are formed through the post-AGB channels \citep{dri1985,heb1986}.
Initial estimates of the death-rate of main-sequence stars and of the birth-rate of white dwarf stars
show broad agreement between the rates although they were still affected by considerable uncertainties \citep[see ][]{dri1985}.
Post-AGB stars are powered by a hydrogen-burning shell in thermal equilibrium interspersed with thermal pulses (helium-shell flash).
The evolutionary tracks of \citet{sch1983} connect the AGB stars to the white dwarfs by invoking a phase of enhanced mass-loss
(or super-wind) that occurs between pulses, i.e., during the hydrogen shell-burning phase. Higher mass-loss rates imply
shorter evolutionary time-scales and, therefore, the assumed rates affect the growth of the carbon/oxygen core
and determine the final mass of the nascent white dwarf.
\citet{wei2000} compared available data to the results of model calculations \citep[e.g.,][]{vas1993} linking the initial main-sequence mass
to the final white dwarf mass. Observations of white dwarfs in stellar systems where evolutionary time-scales
are measurable, i.e., clusters \citep[e.g.,][]{kal2008} or evolved binaries, provide the data and constrain the initial-mass to final-mass
problem \citep{wei2000}. Properties of white dwarfs in spectroscopic and colourimetric studies are parametrized 
by $T_{\rm eff}$ and $\log{g}$ that are transformed into ages ($t_{\rm cool}$) and masses ($M/M_\odot$), hence radii ($R/R_\odot$), using
evolutionary models \citep[e.g.,][]{woo1995,ben1999}.

Extensive colourimetric surveys such as the PG survey helped establish the character of the white
dwarf population such as its local space density and formation rate \citep{gre1980,lie2005} while deeper surveys
such as the Two-Degree Field (2dF) survey allowed to probe the spatial distribution of the white dwarf population and determine 
its Galactic scale-height \citep{ven2002}.
Deep and extensive colourimetric surveys such as the Sloan Digital Sky Survey \citep[SDSS, e.g.,][]{eis2006}
and, potentially, Galaxy Evolution Explorer ({\it GALEX}) all-sky survey combined with kinematical
data \cite[e.g.,][]{sal2003,lep2005} should help retrace the Galactic history, i.e., birth rates and time scales, of the white dwarf population.

We have initiated a program aimed at identifying the white dwarf population in the {\it GALEX} all-sky survey.
As a first instalment, we present a new sample of bright subluminous stars based on 
{\it GALEX} and matched with optical/infrared source catalogues (Section 2).
The sample is particularly suitable for GAIA Calibration \citep{sou2008} and adds
material for surveys of pulsating, hot subdwarfs \citep{bil2002,ost2010}.
We describe the source selection in Section 2.1, the spectroscopic follow-up in Section 2.2, and
the model atmospheres employed in the analysis in Section 2.3. Next,
we present the bright ($V<12$) catalogue in Section 3.1 and a model atmosphere analysis of
the new subdwarfs and white dwarfs in sections 3.2 and 3.3, respectively. We conclude in Section 4.
Photometric ultraviolet, optical, and infrared absolute magnitudes, suitable for the
study of hot white dwarfs in multiwavelength surveys, are listed in Appendix A.

\section{A {\it GALEX/GSC} survey}
\subsection{Source selection}

\begin{figure}
\includegraphics[width=0.98\columnwidth]{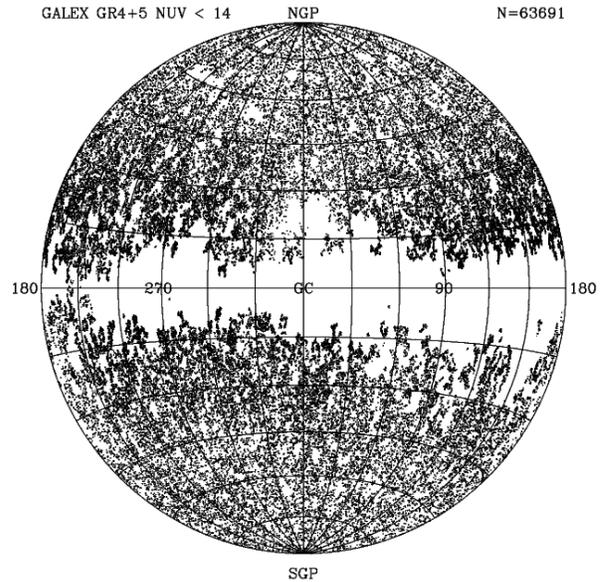}%
\caption{Distribution of the {\it GALEX/GSC} selection in Galactic coordinates.
The selection of 63691 stars with {\it GALEX} 
$N_{\rm UV} <  14$ 
mag is restricted to Galactic latitudes
$|b| \ga 10$-$20^\circ$
and occupies an area of 25000 square degrees.}
\label{fig1}
\end{figure}

We obtained UV photometry from the Galaxy Evolution Explorer
({\it GALEX}) all-sky survey. {\it GALEX} provides photometry in two bands, $F_{\rm UV}$
and $N_{\rm UV}$. The $F_{\rm UV}$ bandpass is $1344 - 1786$ \AA\ (defined at $\ga10$\% peak value) with an effective
wavelength of 1539 \AA, and the $N_{\rm UV}$ bandpass is $1771 - 2831$ \AA\ with an
effective wavelength of 2316 \AA\ \citep[see][]{mor2007}. 
We searched the {\it GALEX} GR4+5 release using the CasJobs batch query services with the following criteria:
\begin{displaymath}
N_{\rm UV} \le 14,\ \ \ N_{\rm UV}\ge 0,\ \ \ F_{\rm UV}\ge 0,
\end{displaymath}
and we extracted the parameters {\it ra}, {\it dec}, {\it nuv\_mag}, {\it nuv\_magerr}, {\it fuv\_mag}, {\it fuv\_magerr}, {\it fov\_radius}, and {\it e\_bv}.
The right ascension and declination ({\it ra}, {\it dec}) supplied by {\it GALEX} are J2000. The $N_{\rm UV}$ and $F_{\rm UV}$ magnitudes ({\it nuv\_mag},
{\it fuv\_mag}) and associated errors ({\it nuv\_magerr}, {\it fuv\_magerr}) are defined in the AB system.
We opted for magnitudes calculated using the SExtractor ``mag\_auto" with a variable elliptical
aperture rather than a fixed aperture.
The distance of the source from the image (or tile) centre ({\it foc\_radius}) is measured in degrees, 
and each tile has a field of view of 1.25$^\circ$ in the $N_{\rm UV}$ 
or 1.27$^\circ$ in the $F_{\rm UV}$. Finally, the total extinction in the line-of-sight ({\it e\_bv}) is based on
the maps of \citet{sch1998}. 
However, reddening corrections were not initially applied because we lack a priori knowledge of the distance and spectral type.

\begin{figure}
\includegraphics[width=0.98\columnwidth]{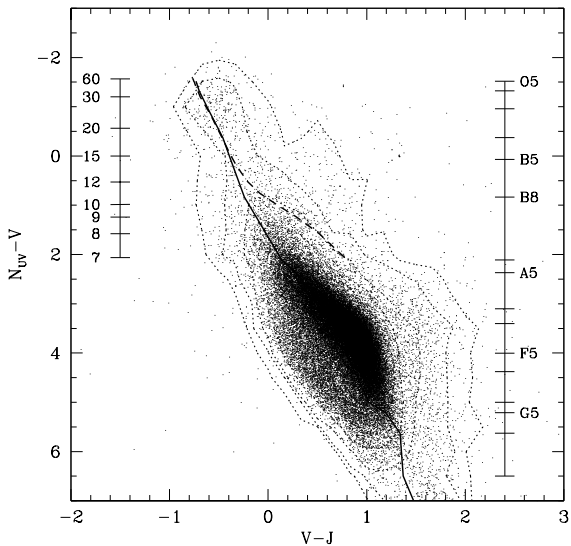}%
\caption{Colour indices ($N_{\rm UV}-V$ versus $V-J$) of 60379 stars from the joint {\it GALEX/GSC} survey. The locations of the main-sequence (full line) and white dwarf cooling sequence (dashed line, $\log{g}=8$) are shown. Contour lines (dotted lines) show the number density of stars (per square magnitude) at 1\%, 3\%, 10\%, and 30\% of the peak density. The corresponding white dwarf temperature scale is shown on the left, and the main sequence spectral
types on  the right. No corrections are applied for interstellar reddening.}
\label{fig2}
\end{figure}

A selection of bright stars with {\it GALEX} suffers potential photometric inaccuracies. First,
the imaging quality degrades toward the edge of the tile \citep{mor2007} altering the point spread function (PSF).
Next, the corrections for photometric nonlinearity are uncertain and depend on the aperture size. 
This effect, compounded with the broadening of the PSF for off-centre sources,
introduces a systematic offset between predicted and measured magnitudes. In practice, we applied
the nonlinearity correction for a 3$\arcmin$ aperture and labelled as potentially inaccurate
the photometry of sources located more than 0.4$^\circ$ from the tile centre.
The requirement that the $F_{\rm UV}$ detection be included could also result in many bright
objects being dropped from the list. A $N_{\rm UV}$-only selection that may potentially triple the membership to the
bright source list will be examined elsewhere.

Entries with coordinates matching within $10^{-3}$ degrees were merged and 
the search resulted in the selection of $N=63691$ individual sources. 
Figure~\ref{fig1} shows
the source locations in Galactic coordinates. The GR4$+$5 release covers $\sim 25000$ square degrees and excludes the Galactic plane. 

\begin{figure*}
\centering
\includegraphics[width=0.98\columnwidth]{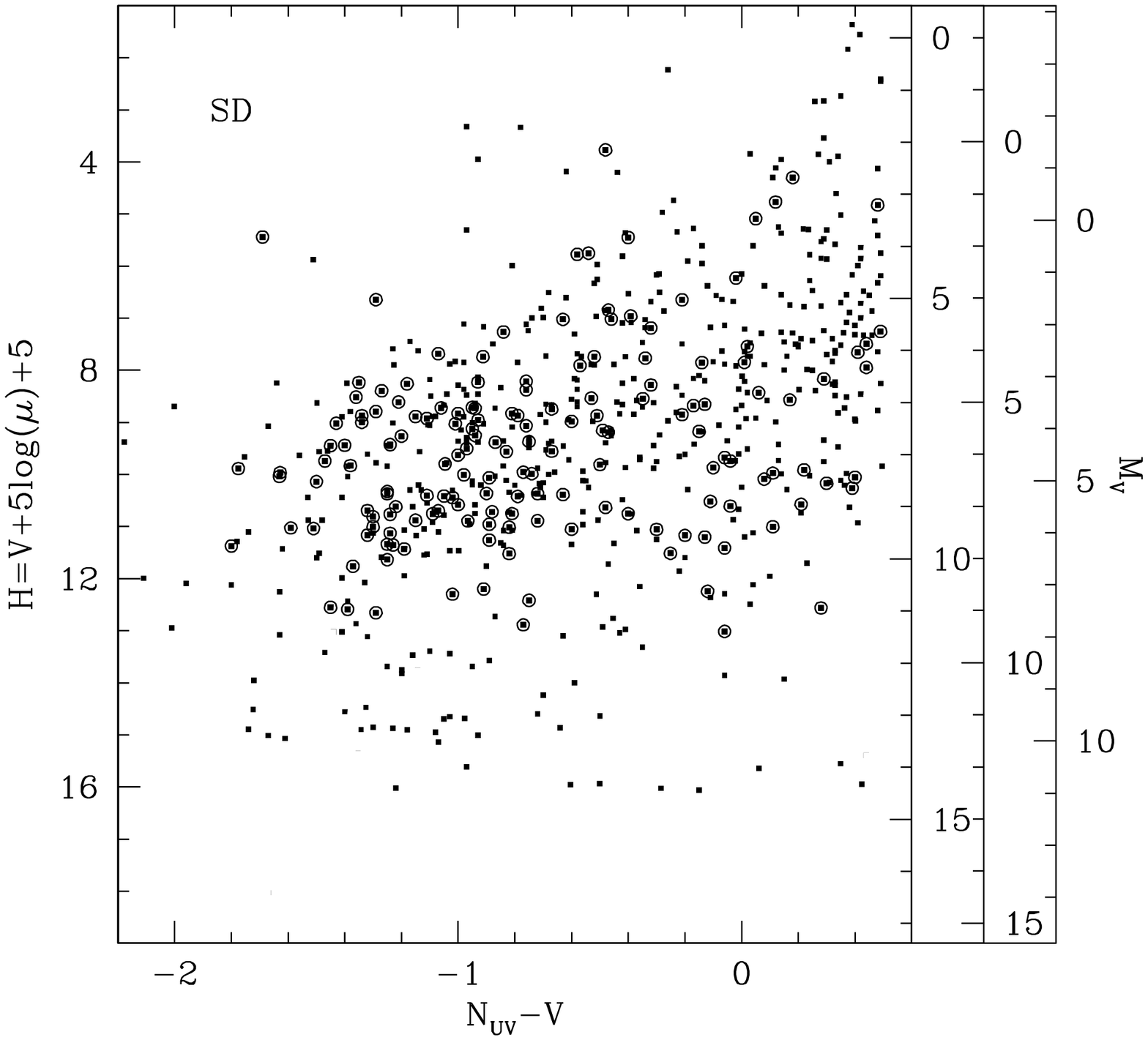}%
\includegraphics[width=0.98\columnwidth]{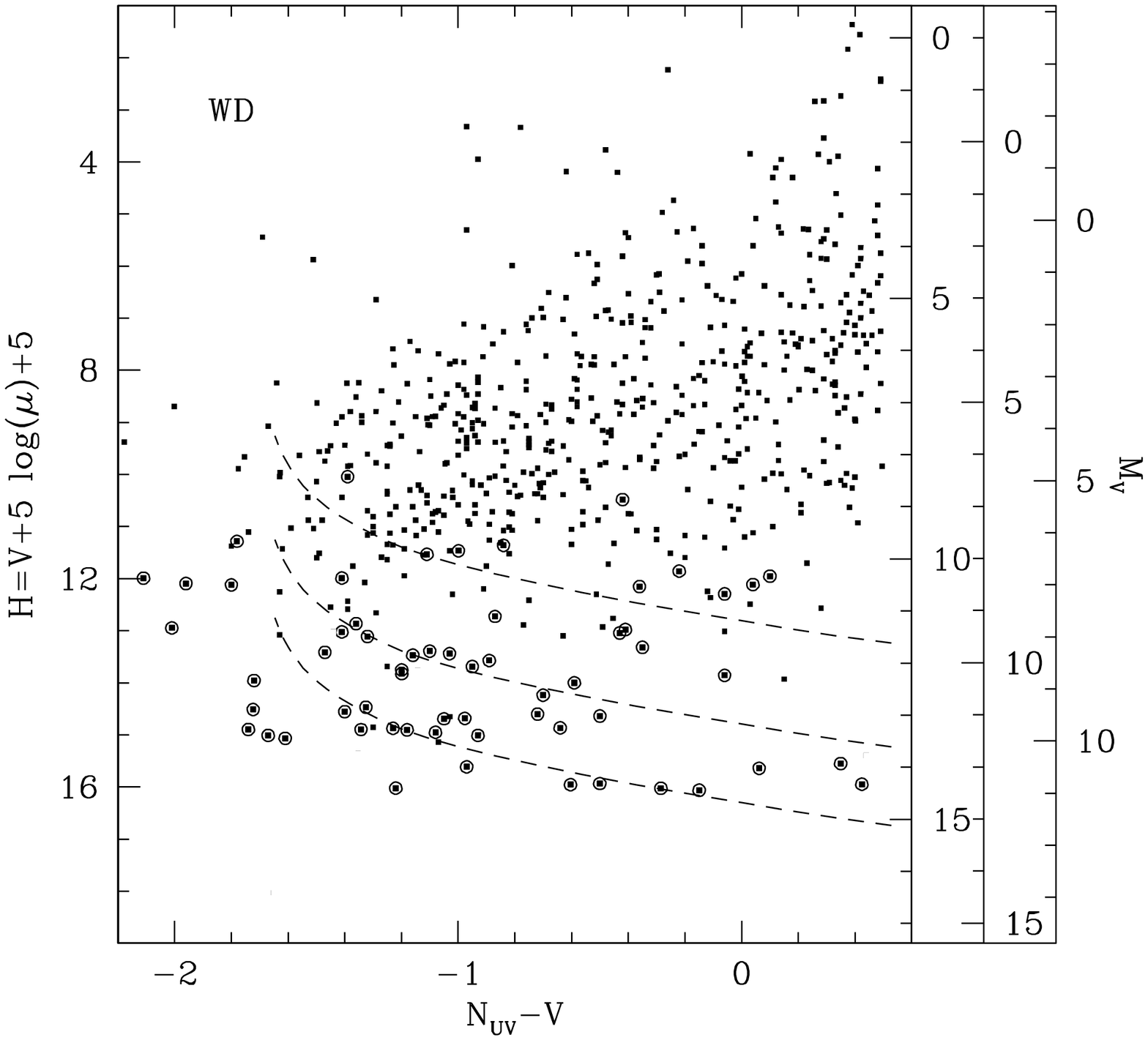}%
\caption{Reduced proper-motion diagram of a sample of $N=694$ UV-excess objects with $N_{\rm UV}-V<0.5$. 
The scales on the right-hand side show the corresponding absolute V magnitudes for transverse velocities of,
from inside to outside, 10, 25, and 50 km s$^{-1}$.
The samples of known hot subdwarfs (left) and white dwarfs (right) are shown with open circles and comprise
159 and 61 objects, respectively. The white dwarf sequence at $0.6\,M_\odot$ (dashed lines) is shown for
tangential velocities of, from top to bottom, 10, 25, and 50 km s$^{-1}$. No corrections are applied for interstellar reddening.
}
\label{fig3}
\end{figure*}

Next, we cross-correlated the
source list with the Guide Star Catalog version 2.3.2 \citep[GSC2.3.2,][]{las2008} using VizieR at the Centre de Donn\'es astronomiques de Strasbourg\footnote{CDS available at http://cdsweb.u-strasbg.fr/CDS.html \citep{och2000}.}.
The closest GSC entry found within a radius of 10 arcseconds of the {\it GALEX} coordinates is thereby associated with the UV source. 
The joint {\it GALEX/GSC} survey comprises $N=60204$ sources, or 96\% of the total, in the 3$\arcsec$ sample and $N=62672$ sources in the 10$\arcsec$ sample.
For each object we recorded both UV magnitudes ($F_{\rm UV}$, $N_{\rm UV}$) and 
we adopted the ``quick-V'' ($V_{\rm 12}$) magnitude which is close to the Johnson $V$ magnitude, $V_{\rm 12}=V-0.15\,(B-V)$ \citep{rus1990}.
Following the same positional criteria we also obtained the infrared photometry from
the Two Micron All Sky Survey \citep[{\it 2MASS};][]{skr2006}, which is available at VizieR on CDS.

Figure~\ref{fig2} shows a $N_{\rm UV}-V$ versus $V-J$ diagram of our {\it GALEX/GSC} selection. 
The data are compared to synthetic main-sequence and white dwarf colours.  
The procedure selects objects with temperatures above 12,000 K, or spectral types earlier than $\sim$B8.

Finally, we built a sample of $N=694$ UV-excess objects by restricting the index $N_{\rm UV}-V<0.5$. 
The choice of the $N_{\rm UV}-V$ index as a primary selection criterion is motivated by the need
to mitigate the effect of binary companions on the selection of white dwarf stars.
Although the 2MASS $J$ magnitude is generally more reliable than the GSC $V_{\rm 12}$ magnitude it is 
potentially contaminated by a putative main-sequence companion to the target white dwarf star (see Section 3.3).

\begin{figure}
\centering
\includegraphics[width=0.50\textwidth]{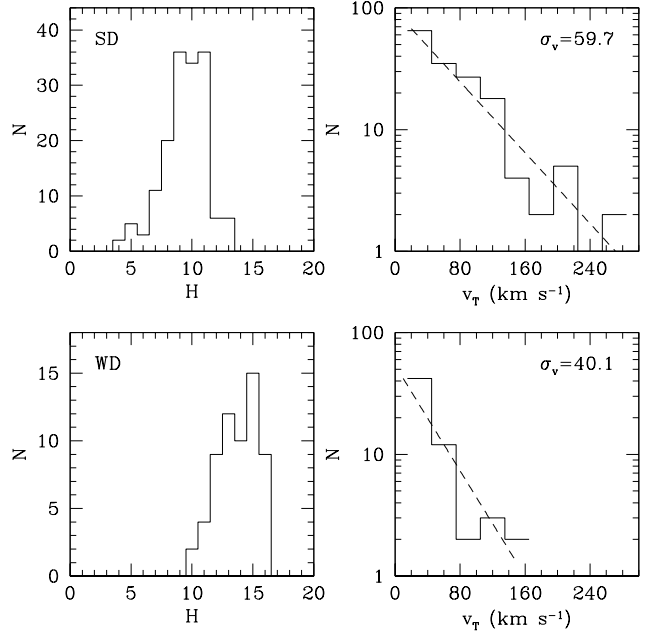} 
\caption{Distribution of $H$ (left) and corresponding transverse velocities $v_T$ (right) for hot subdwarfs (top)
and white dwarfs (bottom). The velocity distribution functions follow $N\propto e^{-v_T/\sigma_v}$, with the best-fit
values given in the upper right corner above the distributions.}
\label{fig4}
\end{figure}

We obtained proper-motion for each object from the Naval Observatory Merged Astrometric Dataset \citep[NOMAD][]{zac2004}
accessed at VizieR. Then, we calculated the reduced proper-motion index $H$:
\begin{displaymath}
H = V+5\log{\mu}+5 = M_V+5\log{v_T}-3.3791,
\end{displaymath}
where $\mu$ is the proper-motion in units of arcsecond yr$^{-1}$, and $v_T$ is the transverse velocity in km s$^{-1}$.
Figure~\ref{fig3} shows the index $H$ versus the colour index $N_{\rm UV}-V$ for the sample of 694 objects with, on the left
panel, the subset of known subdwarfs marked, and, on the right, the known white dwarfs also marked.
A white dwarf sequence at $0.6\,M_\odot$ (see Appendix A) is drawn covering the temperature between $T_{\rm eff}=12000$ and 84000 K
at three representative transverse velocities. Figure~\ref{fig4} shows the corresponding H distribution functions for the
known white dwarfs and subdwarfs.
Hot white dwarfs dominate the number count at $H\ga13$ while hot subdwarfs
are predominant in the range $H\la 12$. 
The H distribution functions are converted into a transverse velocity distribution by calculating the absolute
$V$ magnitude $M_V$ for each object and solving for the velocity.

\citet{gre1974} already suggested that stars on the EHB have similar luminosities. Based on $g\theta^4$ 
values of 2.35 \citep{gre1974}, 2.28 \citep{moe1990a}\footnote{\citet{ede2003} state that the Str\"omgren calibration used by \citet{moe1990a} may have been
inappropriate for sdB stars which may explain their lower mean value for $g\theta^4$. Nonetheless, we include the value in the average and
account for the possible uncertainties.}, 2.64 \citep{saf1994}, and 2.73 from the population
synthesis of \citet{han2003}, we
adopted a mean $\bar{M}_V\approx4.2\pm0.5$ for the 
subdwarfs. We found that the transverse velocities follow a distribution of the form $N\propto e^{-v_T/\sigma_v}$ with 
$\sigma_v \approx 60$ km\,s$^{-1}$. 
Based on a similar sample of hot subdwarf stars, \citet{the1997} concluded that these objects belong
to the old thick Galactic disc. 
Similarly, \citet{alt2004} found that 87\% of their sample of sdB stars belong to the thick disc with the
remainder in the Galactic halo.

Next, we estimated $M_V$ for each white dwarf by adopting a mean mass of $0.6\,M_\odot$ 
and by estimating $T_{\rm eff}$ using the $N_{\rm UV}-V$ colour index. We used the mass-radius relations of \citet{woo1995} and
\citet{ben1999}. The resulting white dwarf velocities 
follow the same relation but with $\sigma_v \approx 40 $ km\,s$^{-1}$.
The kinematics of hot white dwarfs also suggest that they belong to the old disc \citep{sio1988}. 

The apparent deficit in blue ($N_{\rm UV}-V\la-0.5$), luminous ($H\la6$) stars in the upper-left corner of Figure~\ref{fig3} is caused
in part by interstellar extinction affecting more distant stars and in part by the relative rarity of luminous blue stars. 
Using the  $R_V=3.2$ parametrization of \citet{car1989}, we estimate $A_{\rm N_{\rm UV}}/E_{\rm B-V}=7.75$ while
$A_{\rm V}/E_{\rm B-V} = 3.2$. For a hypothetical, distant star with $E_{\rm B-V}=0.2$, the ultraviolet-optical colour
correction would be $(N_{\rm UV}-V)-(N_{\rm UV}-V)_0 = +0.91$. The corresponding shift in colour would displace an intrinsically blue object
in Figure~\ref{fig3} close to a magnitude toward the right.
This effect, not apparent at lower luminosities, helps draw further distinctions between luminous and subluminous stars
that are not affected to the same degree by interstellar extinction.

The information gathered from the sample of known objects is useful in distinguishing the white dwarf
and subdwarf populations based on the observed $H$ distribution.

\subsection{Follow-up optical spectroscopy}

We obtained low dispersion spectra of subluminous candidates using the EFOSC2 (ESO
Faint Object Spectrograph and Camera) attached to the New Technology Telescope
(NTT) at La Silla Observatory, and using the RC Spectrograph (Ritchey-Chretien Focus Spectrograph) 
attached to the 4-m telescope at Kitt Peak National Observatory (KPNO).

At the NTT on UT 2008 October 19-22, 2009 March 2-4, and 2009 August 23-27 we employed the grism \#11 (300 lines per mm) with a
dispersion of $\approx 4.17$ \AA\ per binned pixel ($2\times2$). The slit width was set at $1\arcsec$ resulting 
in a resolution of $\approx 16$ \AA. At the NTT on UT 2010 March 2-4 we employed grism \#7 (600 lines per mm)
with a dispersion of $\approx 1.96$ \AA\ per pixel. The slit width was set at $1\arcsec$ resulting in a resolution
of $\approx 8$ \AA.
Finally, at KPNO on UT 2010 March 23-26 we employed the KPC-10A grating (316 lines per mm) with a dispersion
of 2.75 \AA\ per pixel in first order and centred on 5300 \AA. 
The slit width was set at $1.5\arcsec$ resulting in a spectral resolution of $\approx5.5$ \AA.

Our spectroscopic observations are progressing toward decreasing $H$ values. The sample $H\ge13$ is completed and the 
sample with $H\ge11$ is 80\% completed. We obtained high signal-to-noise spectra suitable for
detailed line profile analysis. This analysis is based on the calculation of model grids and detailed
synthetic hydrogen and helium line profiles.

\subsection{Model atmospheres}

We computed a grid of high-gravity models in local thermodynamic equilibrium (LTE) suitable for the analysis of hydrogen-rich white dwarfs.
The models are in convective/radiative equilibrium and assume plane parallel geometry. The model opacities include hydrogen bound-bound,
bound-free, and free-free absorptions, as well as H$^-$  bound-free and free-free absorptions at lower temperatures. 
The model opacities also include the hydrogen Rayleigh scattering and the electron scattering.
The grid covers a range of parameters suitable for the analysis of hydrogen-rich white dwarfs with $T_{\rm eff}$ from 7000 to 100000 K,
$\log{g}$ from 5.5 to 9.5, and a pure hydrogen composition. Detailed hydrogen line profiles are computed using the tables
of \citet{lem1997}. 
\citet{kaw2006} provide additional details of the model calculations\footnote{Recent improvements in hydrogen line broadening theory taking into account
non-ideal effects \citep{tre2009} could result in upward revisions of our
$T_{\rm eff}$ and $\log{g}$ measurements by up to 10\% and 
0.15 dex, respectively.}.
We describe in Appendix A and employ in Section 3.3 a set of absolute magnitudes computed using these white dwarf models.

\begin{figure*}
\centering
\includegraphics[width=0.99\textwidth]{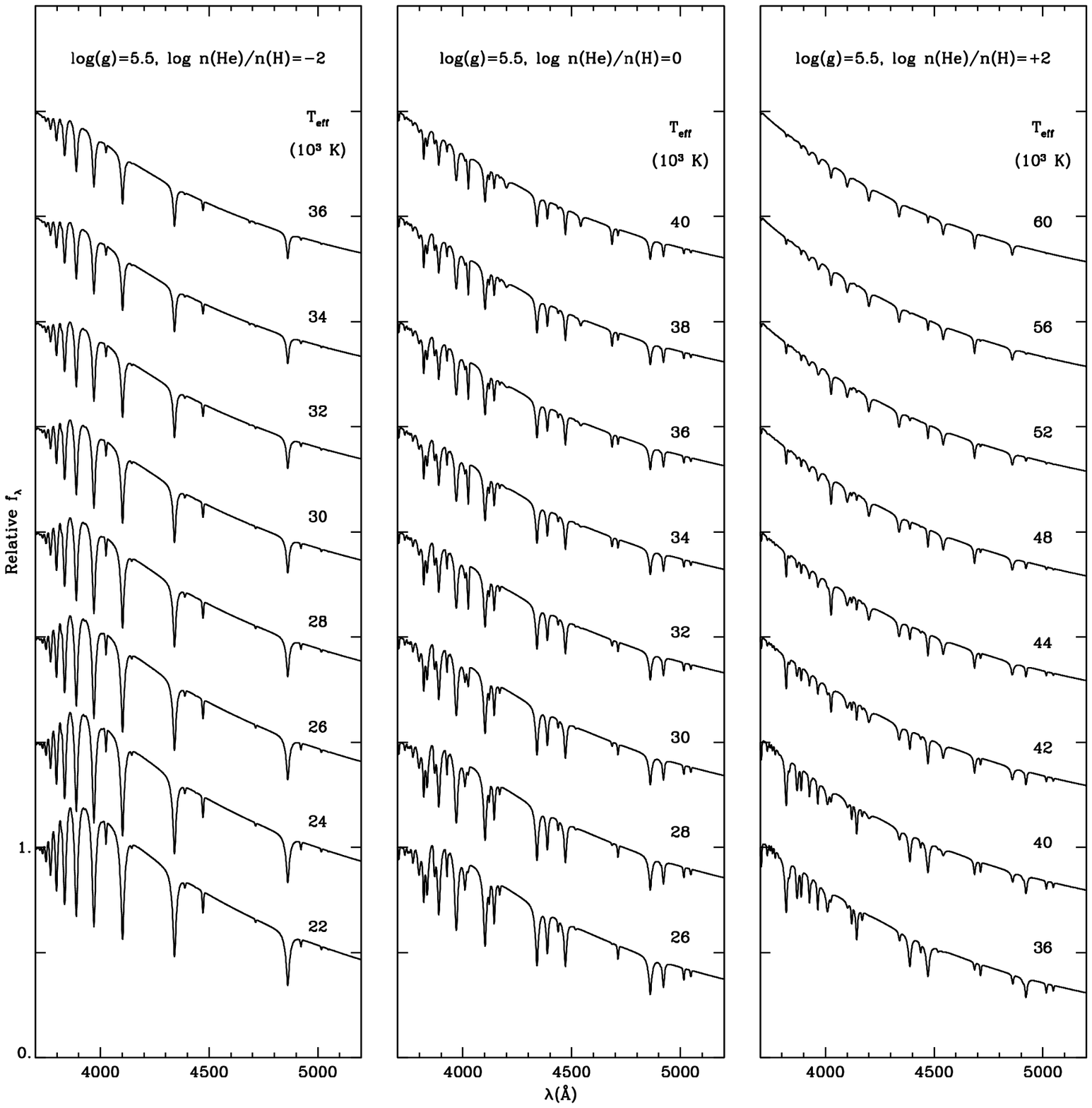}%
\vspace{-0.9cm}
\caption{Model spectra for sdB stars (left panel), He-sdB stars (middle panel), and He-sdO stars (right panel) extracted
from the main model grid.}
\label{fig5}
\end{figure*}

Next, we computed a grid of non-LTE model atmospheres and synthetic spectra covering a range of parameters ($T_{\rm eff}, \log{g}, \log{n({\rm He})/n({\rm H})}$) suitable
for hydrogen and helium-rich subdwarf stars using the codes 
{\sc Tlusty-Synspec} \citep{hub1995,lan1995}. The models are regrouped in three overlapping grids with
grid number 1 suitable for sdB stars covering $T_{\rm eff}$ from 21000 to 35000 in steps of 1000 K,
$\log{g}$ from 4.5 to 6.25 in steps of 0.25 dex, and $\log{n({\rm He})/n({\rm H})}$ from -4 to -1 in steps of 0.5 dex.
Similarly, grid number 2, suitable for sdB and sdO stars, covers $T_{\rm eff}$ from 25000 to 45000, $\log{g}$ from 5.0 to 6.25, and  
$\log{n({\rm He})/n({\rm H})}$ from -3 to 0. Finally, grid number 3, suitable for helium-rich stars, covers $T_{\rm eff}$ from 34000 to 60000
in steps of 2000 K,  $\log{g}$ from 5.0 to 6.25, and $\log{n({\rm He})/n({\rm H})}$ from 0 to 2.
The hydrogen, neutral helium, and ionized helium models comprise 9, 24, and 20 energy levels, respectively.
Figure~\ref{fig5} shows model spectra representative of the sdB and sdO classes.

\begin{figure*}
\includegraphics[width=1.00\textwidth]{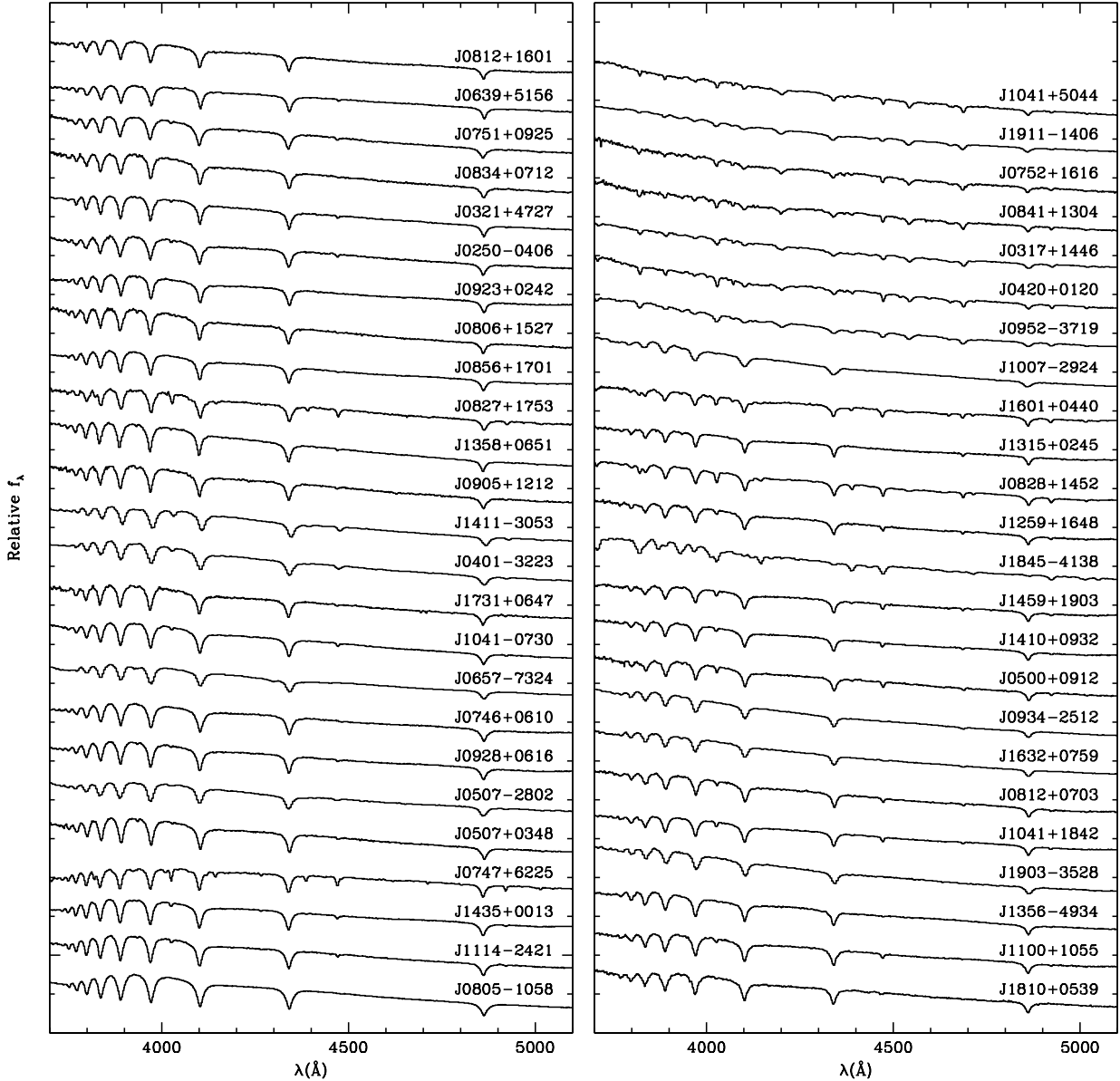}%
\vspace{-0.9cm}
\caption{NTT and KPNO spectra of bright sdB and sdO stars from the {\it GALEX} selection ranked according
to temperatures from the coolest star (lower left) to the hottest star (upper right).}\label{fig6}
\end{figure*}

\section{A Catalogue of subluminous stars}

\subsection{Sample of bright ($V<12$) subluminous candidates}

\begin{table*}
\begin{minipage}{160mm}
\caption{Sample of bright ($V<12$) subluminous candidates.}
\label{tbl-1}
\centering
\renewcommand{\footnoterule}{\vspace*{-15pt}}
\begin{tabular}{rccccccc}
\hline
\hline
  RA (J2000) & Dec (J2000)  & $N_{\rm UV}$ & $V_{\rm 12}$ & $\mu$ \footnote{From NOMAD \citep{zac2004}.} & Name & Type & Telescope \\
             &           & (mag) & (mag) & (mas yr$^{-1}$) & & &    \\
\hline
  0 13 01.0 &   +72 31 19 & 11.97 & 11.49 & $9.3\pm6.3$   & NGC~40          & PNN   & ...  \\
  2 10 22.9 &   +29 27 27 & 12.32 & 11.91 & $6.6\pm0.9$   & uvby98 610242080& B     & 3.58m \\
  2 45 24.7 &   +32 45 29 & 11.45 & 11.33 & $3.6\pm0.8$   & TYC~2329-941-1  & B     & 3.58m \\
  2 45 56.8 &   +11 50 15 & 11.55 & 11.87 & $9.2\pm2.9$   &  ...            & B     & 3.58m \\
  3 15 08.3 &   +14 53 49 & 11.45 & 10.97 & $5.9\pm3.3$   & US~3783,HIP~15137 & B \footnote{The sdB classifications of \citet{mit1998}, \citet{weg1990}, and \citet{wag1988} for these objects are not confirmed.}   & 3.58m \\ 
  3 21 39.8 &   +47 27 17 & 11.91 & 11.70 & $59.5\pm0.7$  & Cl~Melotte~20~488 & sdB \footnote{New subdwarfs in close binaries, see \citet{kaw2010a}.} & 4.0m \\ 
  4 01 05.4 & $-$32 23 48 & 11.03 & 11.20 & $31.3\pm1.7$  & CD-32~1567      & sdB   & 3.58m \\
  4 24 41.2 & $-$20 07 12 & 12.13 & 11.92 & $57.9\pm3.7$  & IM~Eri          & NL    & ... \\ 
  4 47 04.2 & $-$67 06 52 & 11.61 & 11.26 & $<4.3$        & HD~270754       & B1.5Ia \footnote{B1.5 Ia star in the LMC \citep{fea1960} with a B2Ia ultraviolet classification \citep{neu1999}.} & ... \\
  5 05 30.7 &   +52 49 50 & 10.63 & 11.74 & $90.9\pm2.4$  & G191-B2B        &  DA WD \footnote{DA white dwarf WD~0505+527.} & ... \\
  6 06 13.3 & $-$20 21 07 & 11.81 & 11.80 & $16.2\pm4.0$  & CPD-20~1123, Albus~1 & He-sdB \footnote{The blue star Albus 1 \citep{cab2007} was identified as a sdB-He star by \citet{ven2007}.} & ... \\ 
  6 36 46.5 &   +47 00 25 & 11.91 & 11.88 & $2.5\pm1.0$   & ...             & B     & 4.0m \\
  6 39 21.4 &   +33 23 52 & 11.77 & 11.63 & $5.6\pm1.3$   & ...             & B     & 4.0m    \\
  6 39 52.1 &   +51 56 58 & 12.47 & 11.99 & $37.5\pm3.1$  &  ...            & sdB   & 4.0m \\
  6 57 36.8 & $-$73 24 47 & 10.99 & 11.90 & $114.8\pm1.9$ & CPD-73~420      & sdB   & 3.58m  \\
  7 02 22.2 &   +18 40 32 & 12.29 & 11.98 & $<5.0$        & ...             & B     & 3.58m \\
  7 37 25.2 &   +37 14 01 & 10.57 & 10.81 & $6.1\pm1.3$   & KUV~07341+3721,TD1~31205  & B \Large{$^b$} & 4.0m \\ 
  7 47 21.9 &   +62 25 42 & 11.67 & 11.91 & $<9.8$        & FBS~0742+625    & sdB   & 4.0m \\ 
  8 08 37.6 &   +29 08 40 & 11.90 & 11.44 & $12.1\pm1.8$  & TYC 1939-499-1  & A     & 4.0m \\
  8 25 13.5 &   +73 06 38 & 12.27 & 11.78 & $19.7\pm2.8$  & Z~Cam           & DN    & ...   \\
  8 28 32.9 &   +14 52 05 & 11.65 & 11.78 & $23.7\pm1.2$  & TD1~31206,UVO~0825+15 & He-sdB \footnote{Classified as a sdO by \citet{ber1980}.} & 3.58m  \\ 
  8 41 38.5 & $-$07 26 02 & 12.08 & 11.82 & $<5.6$        & TYC 4875-465-1  & B     & 3.58m \\
  9 23 00.2 & $-$67 53 14 & 11.49 & 11.27 & $<11.5$       & TYC 9196-1935-1 & B     & 3.58m \\
 10 00 59.0 &   +02 48 05 & 11.62 & 11.23 & $<3.0$        & TYC~247-190-1   & B     & 3.58m \\
 10 35 53.2 & $-$39 04 14 & 11.32 & 11.04 & $7.9\pm2.8$   & TYC 7710-2503-1 & B     & 3.58m  \\
 10 39 07.7 &   +36 45 33 & 11.23 & 11.25 & $9.9\pm1.2$   & CBS~129         & B \Large{$^b$} & 4.0m  \\ 
 10 44 10.6 &   +48 19 02 & 12.20 & 11.79 & $14.9\pm1.0$  & ...             & A     & 4.0m  \\
 11 05 41.5 & $-$14 04 24 & 11.82 & 11.54 & $7.3\pm3.0$   & EC~11031-1348   & sdB+GV \footnote{\citet{ree2004} noted the composite optical-infrared colours.} & 3.58m  \\ 
 12 11 56.5 & $-$46 55 47 & 11.53 & 11.31 & $12.4\pm1.7$  & ...             & B     & 3.58m  \\
 12 40 51.6 & $-$10 54 13 & 11.26 & 10.89 & $<4.2$        & BD-10 3529      & B     & 3.58m \\  
 12 41 51.8 &   +17 31 19 & 10.67 & 11.64 & $37.4\pm2.6$  & Feige~67,TD1~30996 & sdO   & ... \\ 
 12 45 58.8 & $-$43 05 20 & 10.86 & 10.46 & $49.6\pm2.5$  & CD-42~7878, HD~110942 & B  & 3.58m \\ 
 12 57 52.7 & $-$35 27 18 & 11.26 & 11.55 & $9.8\pm2.9$   & ...             & B     & 3.58m  \\
 13 32 59.7 & $-$41 12 17 & 11.72 & 11.43 & $2.6\pm2.1$   & TYC~7792-881-1  & B     & 3.58m  \\
 13 46 47.2 & $-$07 07 31 & 12.15 & 11.81 & $<5.7$        &  ...            & B     &  3.58m    \\
 13 48 45.0 &   +43 37 58 & 11.32 & 11.35 & $11.6\pm2.0$  & FB~140          & B    & 4.0m  \\ 
 14 11 16.0 & $-$30 53 07 & 11.63 & 11.90 & $9.8\pm2.4$   & CD-30 11223, FAUST~3993 & sdB & 3.58m \\ 
 14 27 08.4 &   +72 57 50 & 11.51 & 11.21 & $<19.0$       & ...             & B     & 4.0m  \\
 15 09 48.5 & $-$38 44 54 & 11.20 & 10.93 & $3.8\pm2.9$   & HD~134199,CPD-38~6054 & B  & 3.58m  \\  
 16 29 32.9 &   +80 16 55 & 11.00 & 10.53 & $<4.9$        &   ...           & B     & 4.0m  \\
 17 02 28.3 &   +63 53 31 & 11.58 & 11.99 & $14.1\pm2.7$  & ...             & B     & 4.0m  \\
 17 23 26.6 &   +46 19 02 & 11.88 & 11.55 & $16.3\pm2.3$  & TYC~3508-387-1  & B     & 4.0m  \\
 17 36 51.3 &   +28 06 35 & 11.53 & 11.44 & $26.8\pm0.8$  & TYC~2084-448-1  & sdB+GV  & 3.58m  \\
 18 01 52.5 &   +37 42 09 & 11.72 & 11.37 & $5.4\pm3.0$   & TYC~3102-1109-1 & B     & 4.0m  \\
 18 15 13.0 &   +39 42 31 & 11.93 & 11.51 & $<2.1$        &  ...            & B     & 4.0m \\
 19 11 09.3 & $-$14 06 54 & 11.77 & 11.77 & $49.6\pm4.2$  & ...             & He-sdO   & 3.58m  \\
 20 47 42.0 &   +08 46 56 & 11.12 & 10.79 & $5.8\pm1.8$   & TYC~1089-1800-1 & B     & 3.58m  \\ 
 20 15 04.8 & $-$40 05 44 & 12.02 & 11.85 & $22.1\pm2.9$  & EC~20117-4014   & sdBV & ...  \\
 23 19 58.4 & $-05$ 09 56 & 11.26 & 11.50 & $12.4\pm5.2$  &  Feige~110      & sdO   &  ... \\
 23 49 47.8 &   +38 44 40 & 11.31 & 11.71 & $5.6\pm0.7$   & FBS~2347+385    & sdB \Large{$^c$} & 2.0m \\ 
\hline
\end{tabular} 
\end{minipage}
\end{table*}

The bright selection potentially harbours new nearby subdwarf or white dwarf stars.
Table~\ref{tbl-1} lists bright UV-selected stars ($V<12$) showing the predominance of massive B stars, followed by hot subdwarfs (sdO, sdB), white dwarfs
(DN$=$dwarf nova, PNN$=$planetary nebula nucleus, DA), and a nova-like star (NL).  
The content of this list resembles and complements the TD-1 catalogue of UV-bright stars. Only about 10\% of UV sources listed by \citet{car1983} 
are subluminous stars, compared to $\approx 30$\% in Table~\ref{tbl-1}. The $N_{\rm UV}$ magnitudes range from 10.5 to 12.5 while $m_{\rm AB}(2300{\rm \AA})\le 11.5$ in the 
TD-1 survey, suggesting, as expected, that subluminous stars should dominate fainter selections such as the present {\it GALEX} selection.

The bright source list includes a single white dwarf, G191-B2B, and 15 hot subdwarfs. 
No new white dwarfs were found in the bright source list. 
Six of the subdwarfs were previously known: the hot sdO stars Feige 67 and Feige 110,
the He-sdB star CPD-20 1123 (Albus~1), the TD-1 discovery UVO~0825+15, the sdB plus GV binary EC~11031$-$1348, and the pulsating sdB EC~20117$-$4014 \citep{odo1997}.
An additional nine new subdwarfs complete the bright source list, including the newly identified binaries GALEX~J0321+4727 and GALEX~J2349+3844 \citep{kaw2010a}.
The sdB GALEX~J0321+4727 is in a 0.26584 d binary with a late-type companion showing a large reflection effect, while the sdB GALEX~J2349+3844
is in a 0.46249 d binary with, most probably, a white dwarf companion.
The brightest new subdwarf in our sample is CD$-$32~1567 with $V=11.2$. 

Samples of bright, nearby white dwarfs are largely based on the Lowell proper-motion survey \citep{gic1980}, listed with
the GD and GR prefixes, or the New Luyten Two-Tenth survey \citep{luy1979,luy1980,sal2003}, listed with the NLTT prefix.
Of the ten known white dwarfs brighter than $V=12$, only RE~2214$-$492 is a relatively recent discovery \citep{hol1993} 
in the ROSAT/Wide Field Camera extreme-UV survey \citep[see ][]{mas1995}.
Based on their expected UV flux output, eight out of the ten known white dwarfs brighter
than $V=12$ were eligible for detection, while the remaining two objects (WD~1142$-$645 and WD~0839$-$327) have low UV flux output. However, 
the fields surrounding WD~1142$-$645, WD~0839$-$327, Procyon~B (WD~0736+053), Sirius~B (WD~0642-166),  
WD~2032+248 (Wolf~1346), and WD~1620$-$391 (CD$-$38 10980) are close to the Galactic plane in areas not covered in the
{\it GALEX} survey. Three of the remaining objects suffered various defects mainly caused by their excessive UV brightness: WD~0413$-$077 (40~Eri~B) is in fact not included in GSC2.3.2 
although it would still have been excluded from our selection 
because of the poor quality of the $N_{\rm UV}$ and $F_{\rm UV}$ photometry; WD~0310$-$688 was also excluded because of the unreliability of the $N_{\rm UV}$ photometry;
RE~2214$-$492 was excluded from our selection because of the lack of $F_{\rm UV}$ photometry. Because it met all the criteria, WD~0505+527 (G191-B2B) is 
the only white dwarf retrieved in the bright source list.

\subsection{Properties of the hot subdwarfs}

Figure~\ref{fig6} shows our spectroscopy of subdwarf stars sorted by effective temperatures. Helium line strengths show large 
variations from one object to the next with the noteworthy cases of J0747+6225 and J0827+1753 which display strong He{\sc i} line spectra.
The spectra of GALEX~J0934$-$2512 and J1632+0752 ($=$PG~1629+081) show peculiar He{\sc i}$\lambda$4471/He{\sc ii}$\lambda$4686 line ratios
that are too weak compared to other objects with similar Balmer line spectra. 
The discrepancy is likely to result in helium abundance inconsistencies. 
Two of our targets showed composite sd+MS spectra and are not shown here (but see Section 3.2.1).
Ten objects show He-rich spectra with the remainder showing H-rich spectra, or a 1:5 number ratio.

\begin{figure}
\centering
\includegraphics[width=0.50\textwidth]{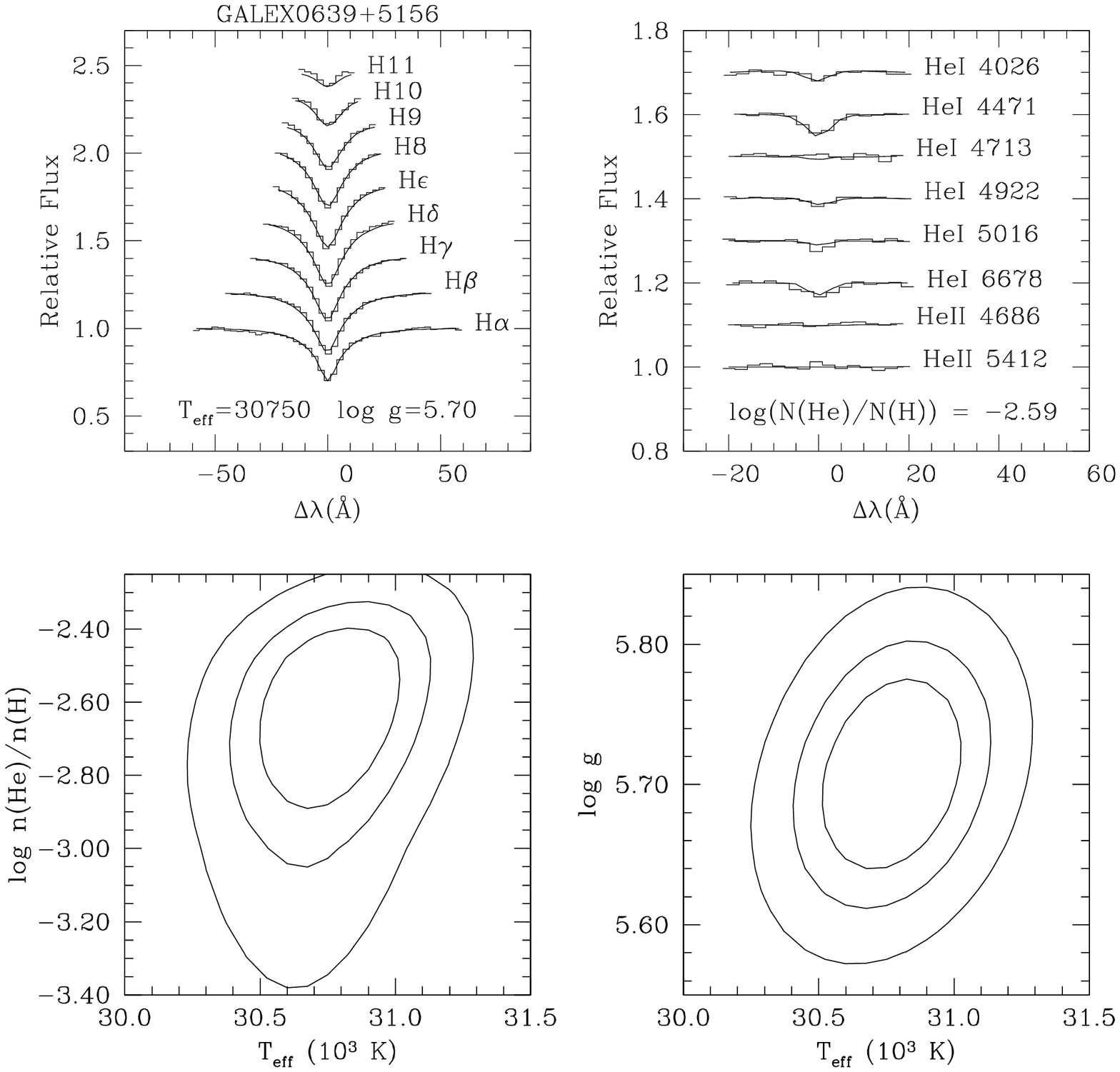}%
\vspace{-0.4cm}
\caption{Model atmosphere analysis of the low-dispersion KPNO spectrum of  GALEX~J0639+5156.  
The top panels show the line profiles and best-fit models, while the lower panels
show the $\chi^2$ contours at 60, 90, and 99\%.}
\label{fig7}
\end{figure}

We fitted the observed spectra from H$\alpha$ (or H$\beta$) to H$_{11}$ with the non-LTE model grid using $\chi^2$ minimization techniques.
The quoted errors are 1-$\sigma$ statistical errors.
Figure~\ref{fig7} shows an example of an analysis of hydrogen/helium line profiles. The hot sdB star GALEX~J0639+5156 is
part of our bright sample ($V\sim12$) and is typical of its class.
Table~\ref{tbl-2} lists the atmospheric parameters of the hot subdwarfs observed in this program.
We listed the results of our analysis for GALEX~J0934$-$2512 and J1632+0752 excluding the He{\sc ii}$\lambda$4686 line.
A detailed analysis of these peculiar subdwarfs will be presented elsewhere after high-dispersion red and blue spectra are obtained. 
A preliminary analysis of the composite sdB+GV spectra is presented in Section 3.2.1. 

The spectral classification follows a simple scheme \citep[see a discussion in][]{heb2009}: spectra
with dominant hydrogen Balmer lines along with weaker He{\sc i} or He{\sc ii} lines are labelled sdB or sdO,
respectively. Spectra with dominant He{\sc ii} lines are labelled He-sdO, whereas spectra
with unusually strong He{\sc i} lines are labelled He-sdB. The newly identified He-sdO GALEX~J1911$-$1406 is a
bright example of its class. 
The low-dispersion spectrum shows strong lines of the He{\sc ii} Pickering series,
He{\sc ii}$\lambda 4686$, He{\sc i}$\lambda 4471$, and a strong C{\sc iii-iv}$\lambda\lambda 4647-4658$ blend,
and N{\sc iii}$\lambda\lambda 4634-4641$ multiplet. All He-sdO stars share these features to varying degrees and
are characterized with $43000 \la T_{\rm eff} \la 52000$ K, $\log{g}\approx 5.6$ and $\log{n({\rm He})/n({\rm H})}\ga 0.7$.
Among their cooler counterparts, such as the He-sdB star GALEX~J0828+1452, some have higher hydrogen abundance than He-sdO stars with weaker
carbon and nitrogen lines. Overall, our subdwarf selection displays abundance diversities often noted in sdB stars \citep{ede2003} and sdO stars \citep{str2007}.

\begin{table*}
\begin{minipage}{150mm}
\caption{Properties of evolved subluminous stars.}
\label{tbl-2}
\centering
\renewcommand{\footnoterule}{\vspace*{-15pt}}
\begin{tabular}{lcllcrl}
\hline
\hline
 \ \ \ \ GALEX~J      & Type & \ \ \ $V$ & \ \ \ \ \ $T_{\rm eff}$ & $\log{g}$ & $\log{n({\rm He})/n({\rm H})}$ & Other names \\
                      &      & (mag)     & \ \ \ \ \ (K)           & (c.g.s.)  &                     &      \\
\hline
025023.8$-$040610  	& sdB    & 13.02    & $28560\pm700$   & $5.56\pm0.15$ & $-2.40_{-0.40}^{+0.25}$ & HE~0247$-$0418, PB~9286\\
031737.9$+$144622\footnote{Str\"omgren photometry from \citet{wes1992}: PG~0314+146 ($y=12.525$), PG~0838+133 ($y=13.653$), PG~0920+029 ($y=14.395$), PG~1257+171 ($y=14.276$), PG~1355+071 ($y=14.307$), PG~1408+098 ($y=14.115$), PG~1432+004 ($y=12.759$), PG~1629+081 ($y=12.779$).}     & He-sdO & 12.67    & $45600\pm2300$  & $5.81\pm0.40$ & $0.82_{-0.27}^{+0.49}$ & PG~0314+146 \\ 
032139.8$+$472716\footnote{Also in the bright ($V<12$) sample.}$^{,}$\footnote{Parameters taken from \citet{kaw2010a}.} & sdB    & 11.79    & $29200\pm300$   & $5.50\pm0.10$ & $-2.60\pm0.10$ & \\
040105.3$-$322348\Large{$^b$} & sdB    & 11.20    & $27000\pm1300$  & $5.36\pm0.19$ & $-1.63\pm0.21$ & CD$-$32~1567 \\
042034.8$+$012041     & He-sdO & 12.52    & $45000\pm800$   & $5.74\pm0.15$ &  $>1.2$  & \\
050018.9$+$091203     & sdB    & 14.63    & $35250\pm570$   & $5.58\pm0.14$ & $-1.25\pm0.12$ &  \\
050720.3$-$280225     & sdB    & 12.39    & $25000\pm2500$  & $4.90\pm0.35$ & $-2.20_{-0.60}^{+0.40}$ & CD$-$28~1974, HE~0505$-$2806 \\  
050735.7$+$034815  	& sdB    & 14.42    & $22050\pm950$   & $5.11\pm0.16$ & $-2.70_{-0.75}^{+0.40}$ &  \\
063952.0$+$515658\Large{$^b$} & sdB    & 11.97    & $30750\pm250$   & $5.70\pm0.08$ & $-2.59_{-0.29}^{+0.19}$  & \\
065736.8$-$732447\Large{$^b$} & sdB    & 11.90    & $26400\pm2000$  & $4.79\pm0.21$ & $<-2.7$ & CPD$-$73~420  \\
074617.1$+$061006     & sdB    & 12.85    & $26100\pm790$   & $5.61\pm0.15$ & $<-3.1$  & \\
074721.9$+$622542\Large{$^b$} & sdB    & 11.91    & $23700\pm800$   & $5.00\pm0.12$ & $-1.00\pm0.10$ & FBS~0742+625 \\
075147.0$+$092526     & sdB    & 14.12    & $29890\pm400$   & $5.65\pm0.11$ & $-2.35\pm0.25$ & \\
075234.2$+$161604     & He-sdO & 14.23    & $48750\pm1450$  & $5.77\pm0.22$ &  $0.96_{-0.26}^{+0.62}$ & \\
080510.9$-$105834     & sdB    & 12.21    & $21200\pm700$   & $5.50\pm0.18$ & $<-2.9$ & \\
080656.7$+$152718     & sdB    & 14.69    & $28400\pm950$   & $5.26\pm0.19$ & $<-2.5$ & \\
081203.8$+$070352     & sdB    & 14.78    & $34900\pm500$   & $5.73\pm0.14$ & $-1.65\pm0.15$ & \\
081233.6$+$160121     & sdB    & 13.57    & $30770\pm500$   & $5.51\pm0.15$ & $<-2.7$ & \\
082734.9$+$175358  	& sdB    & 14.62    & $28200\pm850$   & $5.30\pm0.14$ & $-1.04\pm0.12$ & \\
082832.8$+$145205\Large{$^b$} & He-sdB & 11.78    & $36650\pm650$   & $5.65\pm0.14$ & $-0.50\pm0.08$ & TD1~31206 \\  
083412.3$+$071211     & sdB    & 14.86    & $29000\pm600$   & $5.46\pm0.16$ & $<-2.4$ & \\
084143.8$+$130431\Large{$^a$}     & He-sdO & 13.66    & $47000\pm1800$  & $5.36\pm0.27$ &  $>0.8$ & PG~0838+133 \\
085649.3$+$170115     & sdB    & 13.17    & $28360\pm700$   & $5.30\pm0.15$ & $<-2.6$ & \\
090540.9$+$121228     & sdB    & 14.67    & $27020\pm1000$  & $5.38\pm0.17$ & $-2.70_{-0.60}^{+0.40}$ & PG~0902+124 \\
092308.3$+$024208\Large{$^a$}     & sdB    & 14.48    & $28480\pm800$   & $5.29\pm0.16$ & $<-2.6$ & PG~0920+029 \\
092856.1$+$061634     & sdB    & 14.29    & $25020\pm1100$  & $5.27\pm0.18$ & $-2.74_{-0.60}^{+0.40}$ & PG~0926+065 \\
093448.2$-$251248     & sdO    & 13.03    & $34800\pm800$  & $5.07\pm0.24$ & $-2.77\pm0.62$ \\ 
095256.6$-$371940     & He-sdO & 12.69    & $43000\pm1000$  & $5.44\pm0.30$ & $1.20_{-0.30}^{+0.50}$ &  \\
100752.0$-$292435     & sdB    & 12.87    & $41000\pm950$   & $5.66\pm0.20$ & $<-2.5$   & CD$-$28~7922      \\
104122.8$+$504420     & He-sdO & 15.31    & $51480\pm1500$  & $5.75\pm0.15$ &  $0.88_{-0.16}^{+0.50}$ & PG~1038+510 \\  
104130.4$+$184210     & sdB    & 13.24    & $34100\pm400$   & $5.65\pm0.12$ & $-1.55\pm0.13$ & \\
104148.9$-$073031     & sdB    & 12.14    & $26430\pm850$   & $5.48\pm0.14$ & $-2.30_{-0.27}^{+0.13}$ & \\
110055.9$+$105542     & sdB    & 14.20    & $31550\pm400$   & $5.75\pm0.13$ & $-1.88\pm0.16$ & \\
110541.4$-$140423\Large{$^b$} & sdB+GV & 13.0\footnote{Based on spectral decomposition (see Section 3.2.1).}
                                          & (28000)       & (5.50)        &  ($-$3.0)   &  EC~11031$-$1348, FAUST~2814    \\
111422.0$-$242130     & sdB    & 12.68    & $21560\pm1020$  & $5.02\pm0.20$ & $-2.21\pm0.40$ & EC~11119$-$2405 \\
125941.6$+$164827\Large{$^a$}     & sdB    & 14.37    & $36620\pm530$   & $5.82\pm0.13$ & $-1.47\pm0.13$ & PG~1257+171 \\
131512.4$+$024531     & sdO    & 15.10    & $38980\pm1150$  & $5.14\pm0.19$ & $-1.52_{-0.48}^{+0.42}$ & \\
135629.2$-$493403     & sdB    & 12.30    & $32300\pm530$   & $5.53\pm0.14$ & $-2.70^{+0.42}_{-0.80}$ & CD$-$48 8608 \\
135824.6$+$065137\Large{$^a$}     & sdB    & 14.50    & $28120\pm500$   & $5.45\pm0.10$ & $-2.96\pm0.40$ & PG~1355+071 \\ 
141055.8$+$093256\Large{$^a$}     & sdB    & 14.10    & $36100\pm450$   & $5.68\pm0.11$ & $-1.65\pm0.13$ & PG~1408+098  \\
141115.9$-$305307\Large{$^b$} & sdB    & 11.90    & $26500\pm2700$  & $5.36\pm0.26$ & $-1.34\pm0.24$ & CD$-$30~11223, FAUST~3993 \\ 
143519.8$+$001350\Large{$^a$}     & sdB    & 12.50    & $21650\pm1050$  & $4.99\pm0.18$ & $-2.07\pm0.23$  & PG~1432+004  \\
145928.5$+$190350  	& sdB    & 14.17    & $36100\pm500$   & $5.78\pm0.12$ & $-1.35\pm0.11$ & PG~1457+193 \\
160131.3$+$044027     & He-sdB & 14.56    & $39240\pm600$   & $5.47\pm0.16$ & $-0.50\pm0.11$ & PG~1559+048 \\
163201.4$+$075940\Large{$^a$}     & sdO    & 13.01    & $34700\pm900$  & $5.00\pm0.22$ & $-2.62\pm0.55$ & PG~1629+081 \\  
173153.7$+$064706     & sdB    & 13.74    & $26880\pm1150$  & $5.29\pm0.20$ & $-2.80^{+0.50}_{-1.10}$ & \\
173651.2$+$280635\Large{$^b$} & sdB+GV & 12.6\Large{$^d$} & (28000)         &  (5.50)       &  ($-$3.0)  &       \\
181032.0$+$053909     & sdB    & 13.81    & $31400\pm450$   & $5.85\pm0.14$ & $<-2.5$ & \\
184559.8$-$413827     & He-sdB & 14.63    & $36400\pm3200$  & $5.75\pm0.65$ & $>1.6$   &      \\
190302.5$-$352829     & sdB    & 14.34    & $32900\pm900$   & $5.44\pm0.26$ & $ < -2.5$ &  \\  
191109.3$-$140654\Large{$^b$} & He-sdO & 11.77    & $50000\pm2200$  & $5.56\pm0.32$ & $0.70_{-0.30}^{+0.50}$ & \\
234947.7$+$384440\Large{$^{b,c}$} & sdB    & 11.71    & $28400\pm400$   & $5.40\pm0.30$ & $-3.20\pm0.10$ & FBS~2347+385 \\  
\hline 
\end{tabular} 
\end{minipage}
\end{table*}

\subsubsection{The sdB$+$G2V binaries EC~11031$-$1348 and GALEX~J1736+2806 and four binary candidates}

\begin{figure}
\centering
\includegraphics[width=0.50\textwidth]{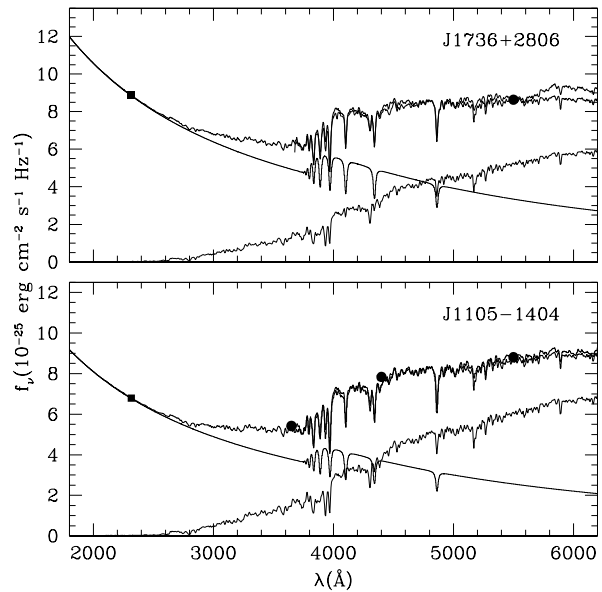}%
\vspace{-0.4cm}
\caption{Spectral energy distribution of the sdB+G2V binaries EC~11031$-$1348 and GALEX~J1736+2806 and spectral decompositions. The hot
subdwarfs are represented with a model at $T_{\rm eff}=28,000$ K and the companion by a G2V template from 
\citet{pic1998}. The $UBV$ photometry for EC~11031$-$1348 is from \citet{kil1997}.
}
\label{fig8}
\end{figure}

\begin{figure}
\centering
\includegraphics[width=0.50\textwidth]{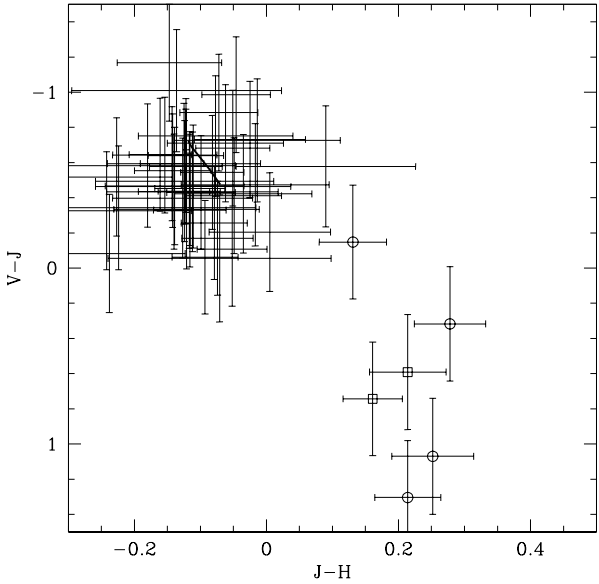} 
\vspace{-0.5cm}
\caption{Optical-infrared $V-J$ versus $J-H$ diagram for new hot subdwarfs (Table~\ref{tbl-2}) showing confirmed or likely
binaries in the lower-right corner. Confirmed binaries are shown with open squares and suspected binaries with
open circles. The companion spectral types range from late F to early G. Most objects cluster near the subdwarf sequence shown
with synthetic colours for models at $20000\le T_{\rm eff}\le 40000$\,K and $\log{g}=5.7$ (thick line). Note
the narrow range of optical-IR colours of hot, single subdwarf stars.}
\label{fig9}
\end{figure}

Companions to hot subdwarf stars are often detected from radial velocity variations
\citep{mor2003,kaw2010a}, while more luminous companions contribute to composite spectra or colours \citep{azn2002,sta2003,ree2004}.
For example, Figure~\ref{fig8} shows a preliminary spectral decomposition of EC~11031$-$1348 and GALEX~J1736+2806. 
Based on infrared excess measurements, \citet{ull1998} found that out of 41 hot subdwarfs, 13 may have G-type companions. 
The fraction is somewhat lower than found by \citet{the1995} who measured a 50\% incidence of binaries in their sample.
Figure~\ref{fig9} shows the location of our own sample of hot subdwarf stars in a $V-J$ versus $J-H$ diagram. Most objects cluster close
to a sequence defined by single, hot
subdwarfs with $V-J, J-H < 0 $. A small group of six objects, two of them analysed here (EC~11031$-$1348, GALEX~J1736+2806),
are located at $V-J, J-H > 0 $ and indicating the presence of a G-type companion in $\approx 1/8$ of our sample.

The subdwarf stars with IR excess are EC~11031$-$1348 \citep{kil1997} which shows composite optical-infrared colours \citep{ree2004}, and
the newly spectroscopically identified sdB+GV GALEXJ~1736+2806. The four other objects with likely F to late G companions
are CD-28~1974 (GALEXJ~0507$-$2802), CPD-73~420 (GALEX~J0657$-$7324),
CD-28~7922 (GALEX~J1007$-$2924) and CD-48~8608 (GALEX~J1356$-$4934).
The two subdwarfs with marked composite spectra (EC~11031$-$1348 and GALEXJ~1736+2806)
are possibly less luminous, hence more easily contaminated in the optical, than the other four objects that show infrared colours typical 
of G-type stars, but that do not show evidence of a companion in blue spectroscopy.

\subsubsection{Overlap with other catalogues of blue stellar objects}

Our selection
recovered thirteen hot subdwarfs from the Palomar-Green survey \citep{gre1986}: the sdB stars PG~0902+124, PG~0920+029, PG~0926+065, PG~1257+171, PG~1335+071,
PG~1408+098, PG~1432+004 \citep{moe1990a,azn2001}, PG~1457+193, PG~1559+048, and PG~1629+081,
and the sdO stars PG~0314+146 \citep{azn2001}, PG~0838+133 \citep{dre1990,the1994}, and PG~1038+510.
A detailed spectroscopic analysis for eight out of ten sdB stars from the PG survey, and two out of three sdO stars from the same survey
is presented here for the first time. \citet{wes1992} list Str\"omgren photometry for eight PG objects from our selection (see Table~\ref{tbl-2}).

Now, we compare the results of our analysis of four PG subdwarfs with the results of previous studies.
 
{\bf PG~0314+146} Using hydrogen models to describe the spectral energy distribution of PG~0314+146 \citet{azn2001} derive a relatively low temperature of 20800K. 
Moreover, \citet{moe1990b} observed narrow hydrogen Balmer lines, 
in contrast to our spectrum which shows broad and shallow He{\sc ii} lines typical of helium-rich sdO stars.
However, \citet{wes1992} noted a discrepancy between their own photometry and that of \citet{moe1990b} and concluded that 
Moehler et al. observed the wrong object.
Our {\it GALEX} target is located at the coordinates listed by Wesemael et al. The {\it International Ultraviolet Explorer} (IUE) low-dispersion 
large-aperture spectrum SWP51740L \citep{azn2001} is peculiar 
with a broad Ly$\alpha$ line and strong C{\sc iv}$\lambda$1550 line, but weaker He{\sc ii}$\lambda$1640 line. The Ly$\alpha$ line strength is inconsistent 
with a photospheric origin. Indeed,
the equivalent width of $W=21$ \AA\ implies a neutral hydrogen column density in the line of sight $n_H=1.9\times10^{18}\ W^2({\rm \AA})=8\times10^{20}$ cm$^{-2}$.
Following the dust-to-gas relation $n_H/E_{\rm B-V}=5.8\times10^{21}$ cm$^{-2}$, the extinction coefficient is predicted to be
$E_{\rm B-V}=0.14$, or $\approx 60$\% of the total extinction in the line-of-sight toward PG~0314+146 \citep[$E_{\rm B-V}=0.2267$, ][]{sch1998}. 
At a Galactic latitude $b=-35.1$ and distance of 300-500 pc (assuming $M_V=4.2\pm0.5$), PG~0314+146 lies 170-290 pc above the plane
and the extinction coefficient is expected to be closer to the total extinction in the line-of-sight.
In summary, PG~0314+146 is a hot He-sdO with a UV flux distribution showing large interstellar extinction.

{\bf PG~0838+133} Our new parameters confirm the analysis of \citet{the1994}, $T_{\rm eff}=55000$ K, $\log{g}=5.8$ and 90\% helium ($\log{n({\rm He})/n({\rm H})}=1$), 
and \citet{dre1990}, $T_{\rm eff}=44000$, $\log{g}=4.8$, $\log{n({\rm He})/n({\rm H})}=1$. We find that the atmosphere is He-rich, as previously
determined, and that our new temperature and surface gravity measurements are consistent with the average of the published values.

{\bf PG~1432+004} Our new measurements agree with the parameters measured by \citet{moe1990a} that are based
on spectroscopy and Str\"omgren photometry ($T_{\rm eff}=22400$\,K, $\log{g}= 5$). However, \citet{azn2001} measured
$T_{\rm eff}=25500\pm700$ K at $\log{g}=5.5$ based on {\it IUE} spectra. The {\it IUE} temperature is significantly higher 
than the optical measurements possibly because of uncertainties in the interstellar reddening index ($E_{\rm B-V}=0.07$)
determined by \citet{azn2001}.

{\bf PG~1629+081} \cite{all1994} and \citet{azn2001} noted that PG~1629+081 has a composite optical-UV spectrum.
However, their temperature measurements, $T_{\rm eff}=32500$ \citep{all1994} and $T_{\rm eff}=26400\pm1150$ K at $\log{g}=5.5$ \citep{azn2001}, 
are lower than expected for a subdwarf showing dominant He{\sc ii}$\lambda$4686 line. We could not verify the presence
of a companion because of the low quality of the 2MASS J photometry.

We recovered two objects from the First Byurakan Survey of blue stellar objects \citep{mic2008}:
FBS~0742+625 classified as B1, and FBS~2347+385 classified as sdB. Our analysis establishes 
sdB classifications for both objects.
We also note from the
Hambourg-ESO survey the objects HE~0247$-$0418 \citep[$=$PB~9286,][]{ber1984} and HE~0505$-$2806 \citep[see][]{fre2006}.
\citet{ber1984} previously classified PB~9286 as a sdB.
Finally, two objects, EC~11031$-$1348 and EC~11119$-$2405, are found in the Edinburgh-Cape survey \citep{kil1997}.
Several objects are also included in the Cape Photographic Durchmusterung (CPD) or Cordoba Durchmusterung (CD) catalogues and
we listed these names in Table~\ref{tbl-2} as long as magnitudes and positions ($r<1\arcmin$) are reasonably matched.

Three hot subdwarfs are also listed in the UV catalogues TD-1 and Far-Ultraviolet Space Telescope \citep[FAUST;][]{bow1995}. The source 
TD1 31206\footnote{The TD-1 catalogue number 32707 used in Simbad is an error.}
\citep{tho1978}, also known as UVO~0825+15 \citep{car1983}, is listed as an sdO by \citet{ber1980} and in the 
catalogue of spectroscopically identified subdwarfs of \citet{kil1988}.
The sources FAUST~2814 and FAUST~3993 are also known as EC~11031$-$1348 \citep{kil1997} 
and CD$-$30~11223, respectively. 

\begin{figure}
\centering
\includegraphics[width=0.50\textwidth]{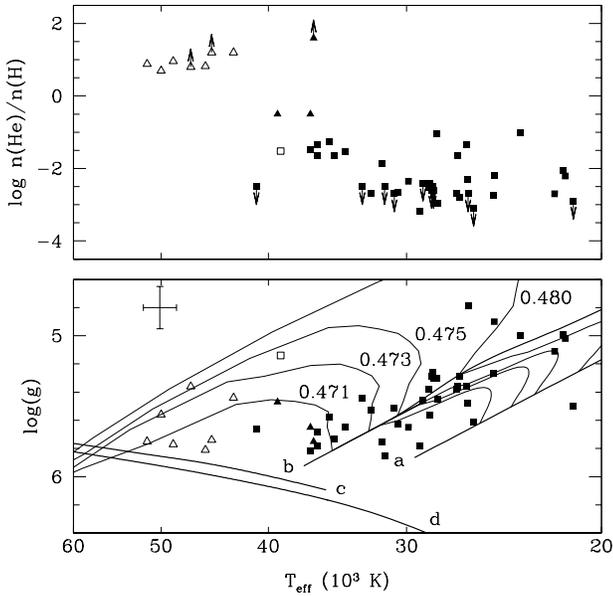} 
\vspace{-0.5cm}
\caption{Parameters of the {\it GALEX} sample of sdO (open square), He-sdO (open triangles), sdB (full squares), and He-sdB stars 
(full triangles). (Top) Helium abundance versus effective temperature measurements showing clear class distinctions. Upper or lower
limits are shown with arrows. (Bottom) Surface gravity versus effective temperature measurements compared to evolutionary
sequences from \citet{dor1993} labelled with the corresponding masses. The ZAEHB and TAEHB are marked with the labels ``a" and
``b", respectively. The HeMS from \citep{pac1971a} and \citet{div1965} are marked with the labels ``c" and ``d", respectively. Typical error bars are shown in the upper left corner.}
\label{fig10}
\end{figure}

\begin{table*}
\begin{minipage}{170mm}
\caption{New white dwarfs selected from colours ($N_{\rm UV}-V<0.5$) and proper motion ($H>13$).}
\label{tbl-3}
\centering
\renewcommand{\footnoterule}{\vspace*{-15pt}}
\begin{tabular}{lccccrllc}
\hline
\hline
\ \ \ \ GALEX~J & $V$   & $N_{\rm UV}-V$  & $N_{\rm UV}-J$ & $J-H$ & $\mu$\ \ \ \    &\ \ \ \ $T_{\rm eff}$ & \ \ \ \ $\log{g}$ & Notes \\ 
                & (mag) & (mag)   & (mag)  & (mag) & (mas yr$^{-1}$) & \ \ \ \ (K)           & \ \ \ \ (cgs)     &       \\
\hline
041051.6+592501\footnote{Extreme UV source identified by \citet{mas1995}.} & 14.69 & $-1.03$  & $-1.18$ & $+0.18$ & $98\pm7$  & 29990$\pm$140   & $7.80\pm0.03$ & 2RE~J0410+592 \\ 
045219.3+251935 & 14.93 & $-1.07$ & $-1.22$  & $-0.07$ & $110\pm3$ & 19620$\pm$370   & $8.23\pm0.06$ & ...  \\
081237.8+173701 & 13.46 & $+0.15$ & $-0.15$  & $-0.08$ & $124\pm3$ & 15260$\pm$160   & $7.82\pm0.04$ & ...  \\
193156.8+011745\footnote{From \citet{ven2010}.} & 14.18 & $-0.63$ & $-1.11$  & $+0.12$ & $61\pm3$  & 20890$\pm$120   & $7.90_{-0.06}^{+0.03}$ & ...  \\
215947.0+293640 & 15.12 & $-1.30$ & $-2.01$  & $-0.07$ & $88\pm2$  & 52850$\pm$1850  & $7.47\pm0.14$ & ...  \\
232358.7+014025\footnote{\citet{har1962}; photometry available from \citet{kil1995}.} & 14.28 & $-1.25$ & $-2.01$  & $-0.13$ & $76\pm4$  & 74300$\pm$3600  & $7.50\pm0.13$ &  PHL~497 \\    
\hline
\end{tabular} 
\end{minipage}
\end{table*}

\subsubsection{Hot subdwarf evolution}

Figure~\ref{fig10} shows the measured parameters in the $\log{g}$ versus $T_{\rm eff}$ plane and the
$\log{n({\rm He})/n({\rm H})}$ versus $T_{\rm eff}$ planes. The temperature and surface gravity measurements are compared to EHB sequences 
at 0.471, 0.473, 0.475, and 0.480 $M_\odot$ \citep{dor1993}. The zero-age extreme horizontal-branch and
the terminal-age extreme horizontal-branch that marks the exhaustion of helium in the core are labelled ZAEHB and TAEHB, respectively.
The location of the helium main sequence (HeMS), covering from 0.33 to $2.0\,M_\odot$ is from \citet{div1965}, \citet{han1971}, and \citet{pac1971a}.
A class distinction is drawn at $\theta = 0.145$. The 10 hot objects ($\theta < 0.145$) are on average more luminous than the 38 cooler objects
($\theta > 0.145$) and are predominantly He-rich. The average value of $\log{g\theta^4}$ is 1.9 in the former group
while it is 2.5 in the latter.
Some He-sdO stars are close to the He-burning main-sequence with masses ranging from 0.75 to $1.0\,M_\odot$ as found by \citet{the1994}
and are possibly more massive than the average sdB star ($\approx 0.5\,M_\odot$).

GALEX~J0804$-$1058 is possibly following an evolutionary path similar to the sdB plus WD binary HD~188112 \citep{heb2003},
and will eventually evolve as an extremely low-mass white dwarf \citep[see][]{kaw2010b}.
A radial velocity study and  search for a potential close companion is under way.

\subsection{Properties of the hot white dwarfs}

We analysed the white dwarf spectra from H$\alpha$ (or H$\beta$) to H$_9$ using the same $\chi^2$ minimization techniques employed in the analysis
of the hot subdwarfs. The results of our analysis using the pure-H model grid are summarized in Table~\ref{tbl-3}.
Close inspection of the spectrum of GALEX~J1931+0117 suggested that the atmosphere of the star is
polluted with heavy elements not normally observed in these objects. \citet{ven2010} presented a detailed
analysis of this peculiar white dwarf.

\begin{figure*}
\centering
\includegraphics[width=0.98\columnwidth]{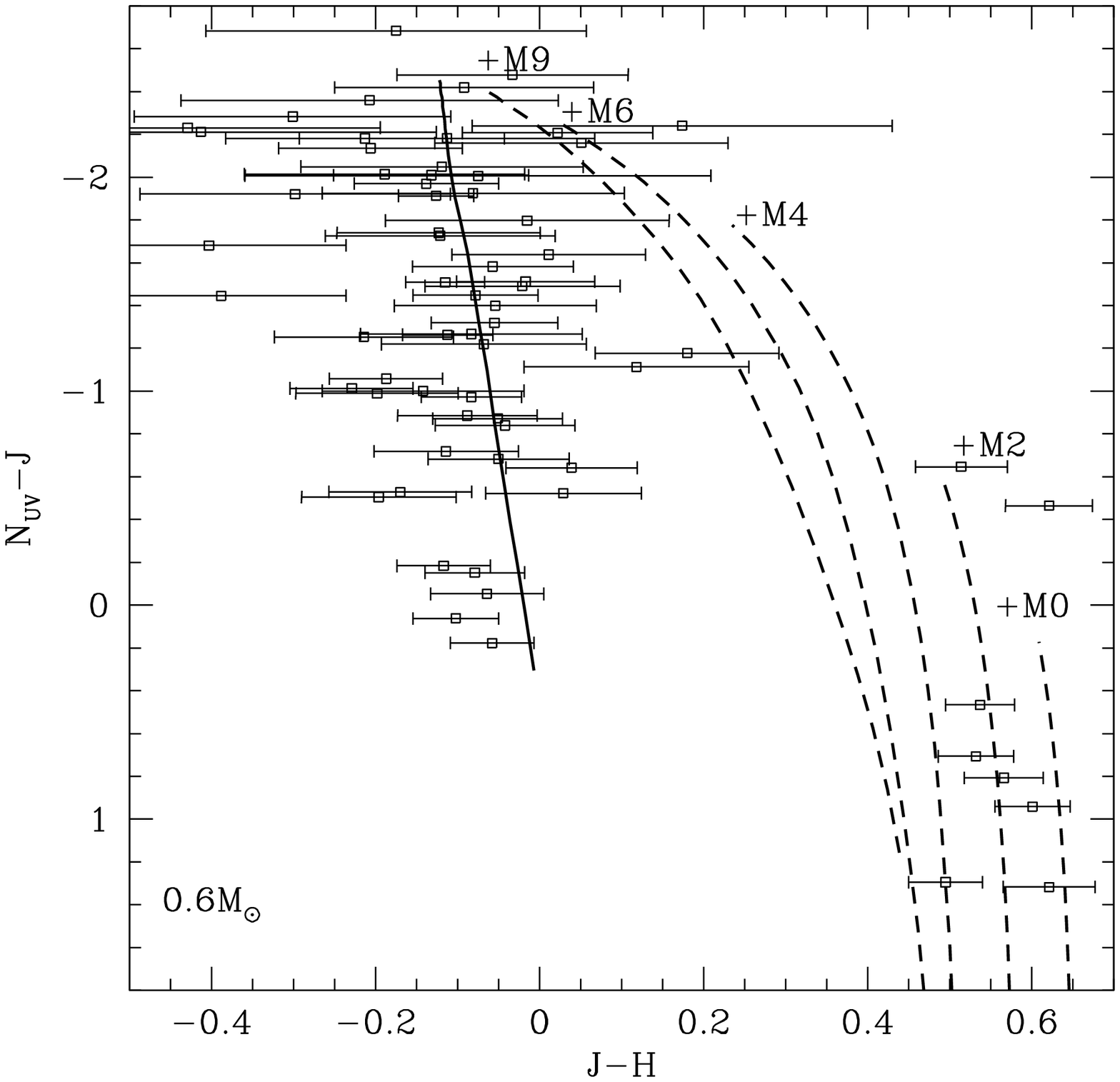} %
\includegraphics[width=0.98\columnwidth]{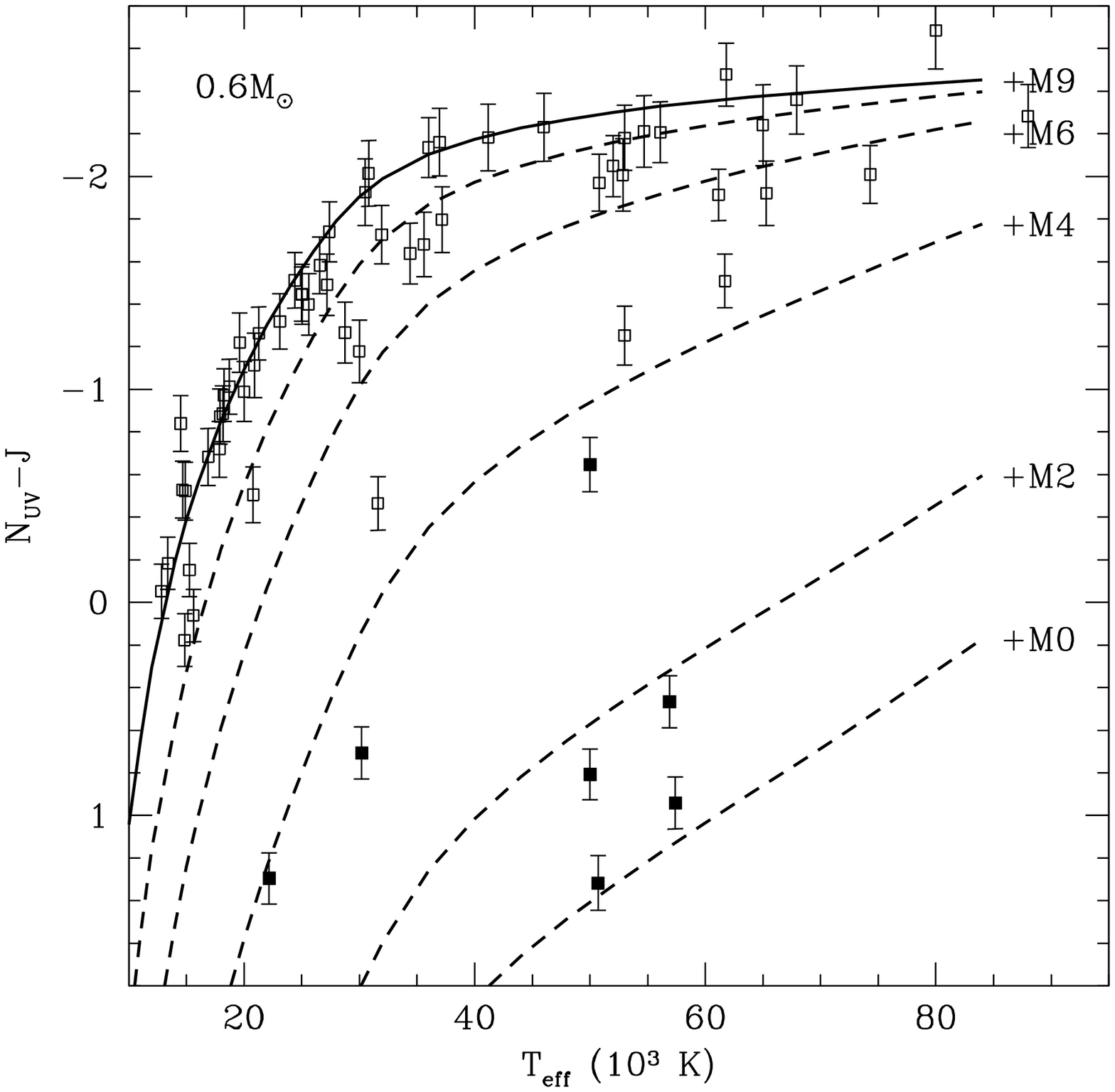} %
\vspace{-0.3cm}
\caption{(Left) UV-infrared $N_{\rm UV}-J$ versus $J-H$ diagram for 6 new hot DA white dwarfs (Table~\ref{tbl-3}) and 50 known 
DA white dwarfs (Table~\ref{tbl-6}) from the present {\it GALEX} selection. Objects in the lower left corner are known white dwarf plus
late-type pairs. The bulk of measurements are close to the hot white dwarf sequence that covers $T_{\rm eff}$ from, top to bottom, 84000 to 12000 K (full line).
Composite colours for DA plus dM2, dM4, dM6, and dM9 are shown with dashed lines. (Right) same as left panel but as a function of effective temperatures.}
\label{fig11}
\end{figure*}

Figure~\ref{fig11} (left) shows hot white dwarfs in a $N_{\rm UV}-J$ versus $J-H$ diagram. Synthetic hot white dwarf colours are listed in Appendix A
and the sample of known white dwarfs is described in Appendix B. Composite
colours were computed by combining absolute infrared magnitudes for M dwarfs \citep{kir1994} with white dwarf absolute magnitudes.
The white dwarf stars with notable IR excess are located in the lower left corner of the diagram. These objects are: PG0205+134, which was
misclassified as a sdO and reclassified as a hot DA \citep{lie2005} with a cool companion \citep{gre1986b,wil2001a,wil2001b},
PG~0824+289 \citep[DA+dC,][]{heb1993}, GD~123 \citep[DA+dM4.5,][]{far2005}, PG~1114+187 \citep[DA+dMe,][]{hil2000}, PG~1123+189
\citep[DA+dM,][]{gre2000}, HZ~43 which is a hot DA with a crowded dM companion \citep{gre1986b}, and the post common-envelope system GD~245 \citep[DA+dMe][]{sch1995}.
Most objects fall along single hot white dwarf colours, but the objects with large infrared excess fall close to the WD+dM2 sequence.

Notable IR excess is also apparent in EUVE~J1847-223 and EUVE~J2124+284 and to a lesser extent in GALEX~J0410+5925 and J1931+0117. 
\citet{ven2010} found a possible link between the infrared excess in GALEX~J1931+0117 and the high heavy element abundance in its
atmosphere. The infrared excess may be the signature of a debris disc being accreted onto the white dwarf surface.
The infrared flux excess in the other three objects remains to be investigated. The case of EUVE~0623$-$376 illustrates potential
problem with the brightest member from the sample. The $N_{\rm UV}-J$ index is affected by inaccurate non-linearity correction of the {\it GALEX}
$N_{\rm UV}$ magnitude measurement.
Figure~\ref{fig11} (right) shows the same sample but as a function of white dwarf effective temperatures. Additional objects, that lay hidden in the
main white dwarf cooling sequence, show a $N_{\rm UV}-J$ index apparently contaminated by late M dwarf companion (dM4-dM6). 
The majority of the known DA white dwarfs closely follow the predicted UV-infrared colour index.
About 10\% of the white dwarfs have early M dwarf companions (M0-M4) with possibly
another 10\% having late M dwarf companions (M4-M9). Early-type companions are only detectable in far and extreme ultraviolet
selections such as that of TD-1 \citep{lan1996} or {\it EUVE} \citep{ven1998}.

\section{Summary and conclusions}

We presented a new selection of $\sim 700$ subluminous stars based on the {\it GALEX} all-sky survey
and UV-optical-IR colour indices built using the GSC2.3.2 $V_{\rm 12}$ and 2MASS J and H magnitudes.
A reduced proper-motion diagram allowed to segregate new white dwarf and subdwarf candidates and
spectroscopic follow-up observations uncovered 48 previously unknown or poorly studied hot subdwarf stars, and six new
white dwarf stars. Two of the new subdwarfs have been found to be in short-period binaries \citep{kaw2010a}
while one of the white dwarfs is a rare high-metallicity DAZ white dwarf \citep{ven2010}.

The content of the bright source list ($V\le12$) is typical of bright UV surveys such as the TD-1 survey. The list
contains primarily blue stars, but also 15 hot subdwarf stars. Nine of the bright subdwarfs were previously unstudied while
the sdB star GALEX~J0639+5156 and the He-sdO star GALEX~J1911$-$1406 were previously uncatalogued.
The bright subdwarf stars are part of a larger list of 52 objects, with 48 of them analysed here for the first time.
We found that six of the hot subdwarfs show composite UV-IR colours indicative of F to K type companions, with two
of them showing composite sdB plus GV optical spectra. The incidence of binaries in our sample is low ($\sim12$\%).
\citet{the1995} and \citet{ull1998} found a higher fraction of binaries in their mixed sample of hot subdwarfs ($>$20\%).
\citet{wil2001a,wil2001b} also conclude based on IR and optical colour criteria that $\ga 60$\% of sdO stars have
cooler companions, while \citet{jef1998} and \citet{ree2004} found a $\sim$20\% incidence of binaries among samples of
hot sdB stars. Relatively inaccurate $V-J$ and $J-H$ colour indices are possibly responsible for the lower
incidence of binaries in our study.

The stellar parameters $T_{\rm eff}$ and $\log{g}$ locate the sample of hot subdwarfs along post-EHB evolutionary 
tracks from the zero-age to the He main-sequence.
The abundance of helium in the sample of sdB stars does not correlate with effective temperature $\la 30000$ K,
but a shallow trend is observed at $30000\la T_{\rm eff}\la 40000$ K \citep[see][]{ede2003}.
A clear break in the helium abundance at $T_{\rm eff}\approx40000$ K separates He-sdO stars from the other classes. The He-sdB GALEX~J1845$-$4138
is an outstanding case among hot sdB stars. Its helium abundance is comparable to He-sdO stars but with a temperature
$\sim 4000$ K cooler.

Although we were able to report the discovery of several new hot subdwarfs, the bright source list does not contain new white dwarfs.
The six new white dwarfs uncovered in this study were selected amongst fainter objects with higher proper-motions ($H>13$). 
New white dwarfs may yet be uncovered amongst unobserved candidates in the range $11<H<13$.

Future work on this program will involve a detailed abundance study, in particular for carbon and nitrogen
that offer clues to the origin of subdwarfs \citep[e.g.,][]{lan2004,ahm2007}. The N{\sc iii}/C{\sc iii}$\lambda\lambda$4634-4651 blend is seen in 
several He-sdO stars. High-dispersion optical and ultraviolet spectroscopy would also enable abundance measurements in sdB stars.
For example, \citet{ohl2000} measured a metallicity of $0.1\times$ solar in the sdB star PG~0749+658:
A more complete abundance pattern would also contribute in improving our
$T_{\rm eff}$/$\log{g}$ measurements that are based on H/He model atmospheres \citep[see][]{ede2003}.

Spectroscopic observations and model atmosphere analyses of a second list of objects 
based on {\it GALEX} GR6 $N_{\rm UV}$ photometry will be reported in a forthcoming paper.

\vspace{-0.3cm}
\section*{Acknowledgments}

S.V. and A.K. acknowledge support from the Grant Agency of the Academy of Sciences of the Czech Republic (IAA~300030908, IAA~301630901)
and from the Grant Agency of the Czech Republic (GA \v{C}R P209/10/0967).
A.K. also acknowledges support from the Centre for Theoretical Astrophysics (LC06014). We thank 
C. Latham and E. Snape for their assistance with the initial catalogue selection, and R.
{\O}stensen for helpful comments on the paper.

\appendix
\section{Ultraviolet, visual, and infrared magnitudes for white dwarfs}

\begin{figure*}
\centering
\includegraphics[width=0.98\columnwidth]{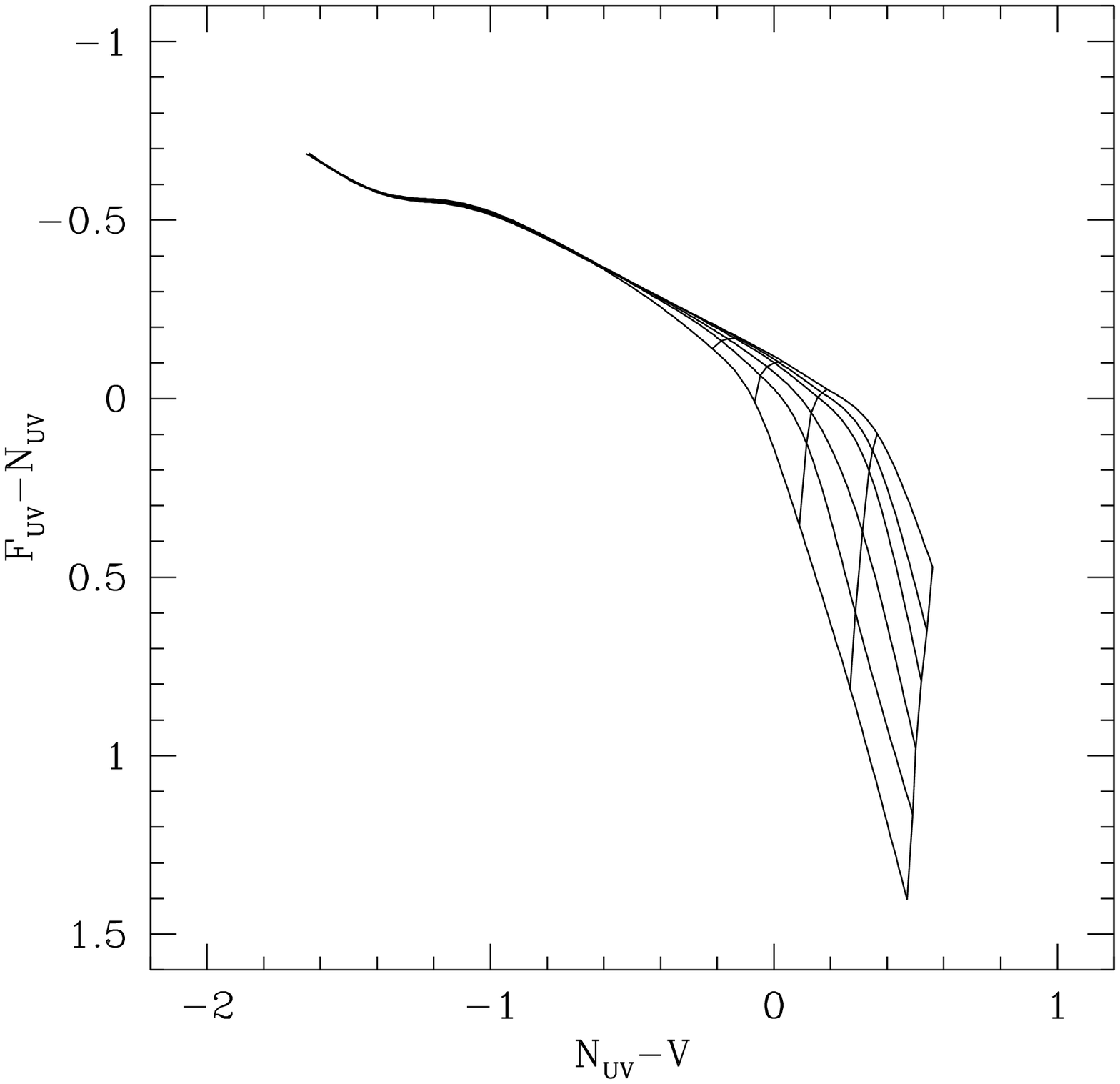}%
\includegraphics[width=0.98\columnwidth]{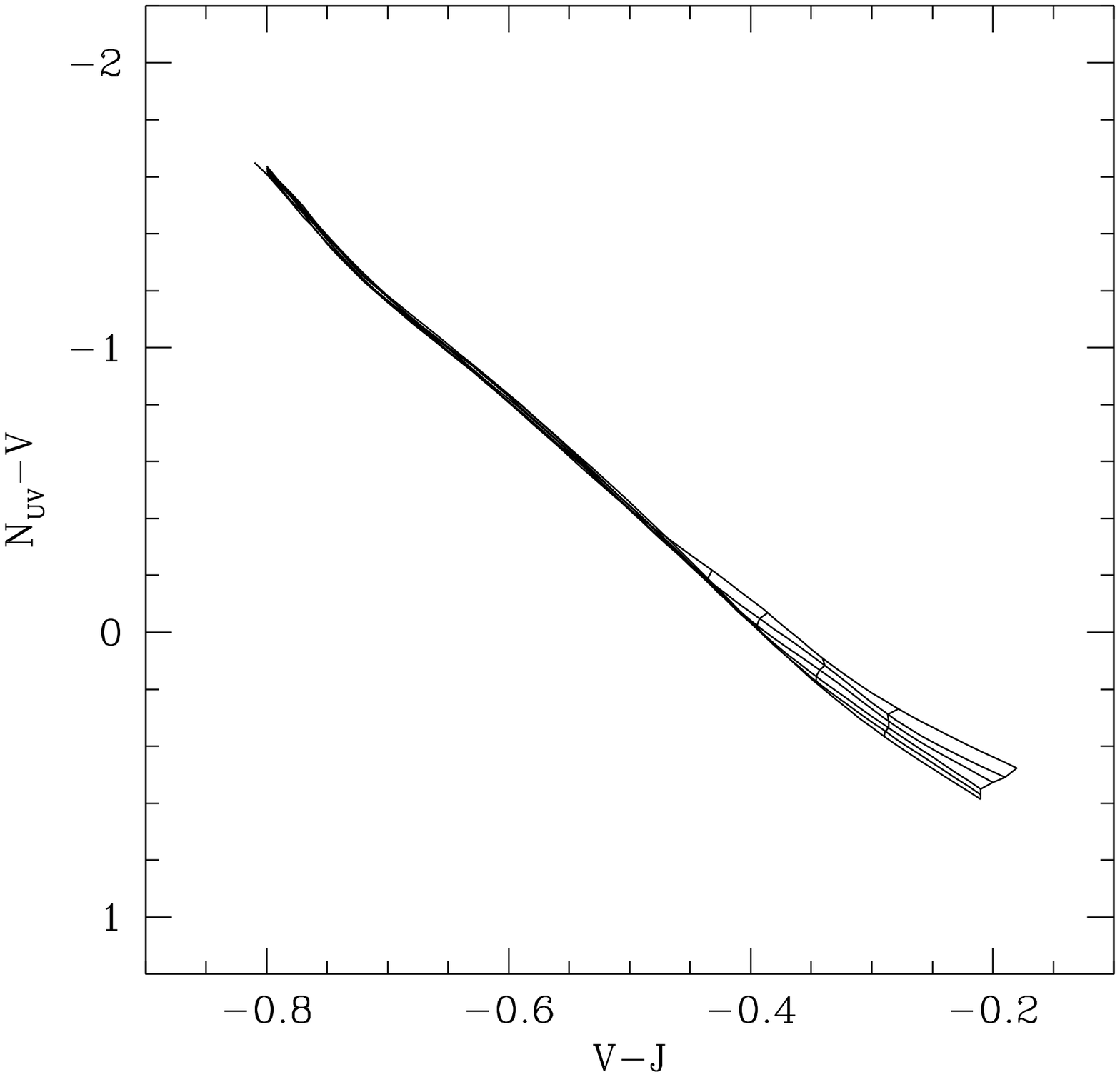}%
\caption{(Left) $F_{\rm UV}-N_{\rm UV}$ versus $N_{\rm UV}-V$ colour indices for white dwarfs
at, from top to bottom, 0.4, 0.5, 0.6, 0.8, 1.0, and 1.2 \,$M_\odot$ and $12000\le T_{\rm eff}\le84000$ K, and (right) 
$N_{\rm UV}-V$ versus $V-J$ for the same masses and temperatures.
}
\label{fig12}
\end{figure*}

Table~\ref{tbl-4} lists absolute {\it GALEX} $N_{\rm UV}$ and $F_{\rm UV}$ magnitudes,
Johnson $V$ magnitude, and the 2MASS $J$ and $H$ magnitudes for white dwarfs of 0.4, 0.5, 0.6, 0.8, 1.0 and 1.2\,$M_\odot$.
The absolute calibration employs the mass-radius relations of \cite{woo1995} and \cite{ben1999}.
We calculated the absolute ultraviolet magnitudes $N_{\rm UV}$ and $F_{\rm UV}$ in the AB system and using the 
post-launch response curves\footnote{Available at {\tt http://galexgi.gsfc.nasa.gov/docs/galex/\\ Documents/PostLaunchResponseCurveData.html}.}.
Next, we calculated the Johnson $V$ and 2MASS magnitudes using the response curves and zero-points of \citet{bes1990}
and \citet{coh2003}, respectively.
Figure~\ref{fig12} shows colour indices for all masses and between 12000 and 84000 K. The  $F_{\rm UV}-N_{\rm UV}$ index
is sensitive to surface gravity, hence mass, at lower temperatures. With increasing surface gravity, the extended Ly$\alpha$ line wing and 
the Lyman satellites act to depress the continuum in the $F_{\rm UV}$ band.
The synthetic colours are computed with pure hydrogen models and do not take into account
the effect of metallicity. The colours should also be corrected for the effect of interstellar
reddening as needed.

\begin{table*}
\begin{minipage}{160mm}
\caption{Absolute ultraviolet, optical, and infrared magnitudes as a function of white dwarf effective temperatures and masses.}
\label{tbl-4}
\centering
\renewcommand{\footnoterule}{\vspace*{-15pt}}
\begin{tabular}{crrrrrccrrrrr}
\hline
\hline
   $T_{\rm eff}$ & $F_{\rm UV}$ &  $N_{\rm UV}$  &   $V$  &  $J$  &  $H$   &&  $T_{\rm eff}$ & $F_{\rm UV}$ &  $N_{\rm UV}$  &   $V$  &  $J$  &  $H$\\
   ($10^3$ K)    &(mag) &(mag) &(mag) &(mag) &(mag) & & ($10^3$ K) &(mag) &(mag) &(mag) &(mag) &(mag) \\           
\cline{1-6}  \cline{8-13}
    \multicolumn{6}{c}{$0.4\,M_\odot$}  & & \multicolumn{6}{c}{$0.5\,M_\odot$} \\
\cline{1-6}  \cline{8-13}
  10.0 &15.103 &12.631 &11.622 &11.574 &11.530 &&   10.0 &15.538 &12.912 &11.921 &11.862 &11.813    \\
  11.0 &13.381 &12.067 &11.289 &11.405 &11.401 &&   11.0 &13.819 &12.353 &11.591 &11.699 &11.688    \\
  12.0 &12.127 &11.644 &11.079 &11.297 &11.310 &&   12.0 &12.566 &11.910 &11.368 &11.588 &11.599    \\
  13.0 &11.372 &11.273 &10.908 &11.198 &11.221 &&   13.0 &11.710 &11.565 &11.217 &11.506 &11.526    \\
  14.0 &10.902 &10.928 &10.739 &11.082 &11.115 &&   14.0 &11.210 &11.226 &11.055 &11.401 &11.431    \\
  15.0 &10.526 &10.629 &10.595 &10.978 &11.018 &&   15.0 &10.827 &10.929 &10.912 &11.300 &11.338    \\
  16.0 &10.192 &10.356 &10.461 &10.877 &10.924 &&   16.0 &10.501 &10.668 &10.789 &11.210 &11.255    \\
  18.0 & 9.607 & 9.867 &10.210 &10.682 &10.740 &&   18.0 & 9.937 &10.203 &10.560 &11.038 &11.095    \\
  20.0 & 9.104 & 9.443 & 9.982 &10.503 &10.570 &&   20.0 & 9.455 & 9.801 &10.353 &10.880 &10.946    \\
  22.0 & 8.658 & 9.065 & 9.770 &10.335 &10.410 &&   22.0 & 9.029 & 9.443 &10.161 &10.730 &10.804    \\
  24.0 & 8.246 & 8.711 & 9.561 &10.165 &10.248 &&   24.0 & 8.643 & 9.114 & 9.976 &10.584 &10.666    \\
  26.0 & 7.869 & 8.378 & 9.357 & 9.998 &10.088 &&   26.0 & 8.292 & 8.806 & 9.795 &10.441 &10.530    \\
  28.0 & 7.522 & 8.058 & 9.151 & 9.825 & 9.923 &&   28.0 & 7.971 & 8.511 & 9.612 &10.291 &10.388    \\
  30.0 & 7.210 & 7.758 & 8.944 & 9.645 & 9.749 &&   30.0 & 7.687 & 8.237 & 9.430 &10.134 &10.238    \\
  32.0 & 6.940 & 7.493 & 8.751 & 9.470 & 9.579 &&   32.0 & 7.442 & 7.997 & 9.260 & 9.982 &10.089    \\
  36.0 & 6.505 & 7.071 & 8.424 & 9.165 & 9.277 &&   36.0 & 7.054 & 7.622 & 8.978 & 9.721 & 9.833    \\
  40.0 & 6.166 & 6.748 & 8.162 & 8.916 & 9.031 &&   40.0 & 6.757 & 7.341 & 8.757 & 9.513 & 9.627    \\
  44.0 & 5.879 & 6.476 & 7.934 & 8.697 & 8.813 &&   44.0 & 6.502 & 7.101 & 8.561 & 9.325 & 9.441    \\
  48.0 & 5.626 & 6.236 & 7.729 & 8.498 & 8.615 &&   48.0 & 6.272 & 6.885 & 8.378 & 9.149 & 9.266    \\
  52.0 & 5.388 & 6.011 & 7.531 & 8.307 & 8.424 &&   52.0 & 6.051 & 6.676 & 8.197 & 8.974 & 9.091    \\
  56.0 & 5.159 & 5.793 & 7.337 & 8.117 & 8.236 &&   56.0 & 5.841 & 6.476 & 8.021 & 8.802 & 8.920    \\
  60.0 & 4.933 & 5.576 & 7.141 & 7.925 & 8.044 &&   60.0 & 5.629 & 6.274 & 7.838 & 8.624 & 8.743    \\
  64.0 & 4.700 & 5.352 & 6.934 & 7.723 & 7.842 &&   64.0 & 5.426 & 6.079 & 7.661 & 8.451 & 8.570    \\
  68.0 & 4.455 & 5.116 & 6.715 & 7.507 & 7.627 &&   68.0 & 5.220 & 5.882 & 7.480 & 8.272 & 8.392    \\
  72.0 & 4.193 & 4.861 & 6.475 & 7.270 & 7.391 &&   72.0 & 5.018 & 5.687 & 7.299 & 8.095 & 8.216    \\
  76.0 & 3.912 & 4.587 & 6.215 & 7.013 & 7.135 &&   76.0 & 4.806 & 5.482 & 7.107 & 7.905 & 8.026    \\
  80.0 & 3.593 & 4.274 & 5.914 & 6.715 & 6.838 &&   80.0 & 4.594 & 5.276 & 6.912 & 7.714 & 7.835    \\
  84.0 & 3.175 & 3.862 & 5.514 & 6.318 & 6.441 &&   84.0 & 4.374 & 5.062 & 6.709 & 7.513 & 7.636    \\
%
\cline{1-6}  \cline{8-13}
    \multicolumn{6}{c}{$0.6\,M_\odot$}  & & \multicolumn{6}{c}{$0.8\,M_\odot$} \\
\cline{1-6}  \cline{8-13}
  10.0 &15.898 &13.158 &12.182 &12.115 &12.062 &&  10.0 &16.545 &13.633 &12.681 &12.600 &12.540   \\
  11.0 &14.195 &12.603 &11.854 &11.953 &11.938 &&  11.0 &14.902 &13.077 &12.349 &12.435 &12.411   \\
  12.0 &12.952 &12.142 &11.617 &11.836 &11.843 &&  12.0 &13.606 &12.606 &12.100 &12.309 &12.309   \\
  13.0 &12.016 &11.813 &11.477 &11.763 &11.780 &&  13.0 &12.634 &12.259 &11.946 &12.232 &12.244   \\
  14.0 &11.478 &11.482 &11.327 &11.673 &11.700 &&  14.0 &11.990 &11.950 &11.818 &12.161 &12.182   \\
  15.0 &11.088 &11.188 &11.186 &11.577 &11.613 &&  15.0 &11.566 &11.655 &11.679 &12.074 &12.105   \\
  16.0 &10.758 &10.926 &11.061 &11.487 &11.530 &&  16.0 &11.231 &11.398 &11.560 &11.992 &12.031   \\
  18.0 &10.202 &10.473 &10.843 &11.325 &11.380 &&  18.0 &10.676 &10.953 &11.346 &11.836 &11.887   \\
  20.0 & 9.729 &10.081 &10.645 &11.176 &11.240 &&  20.0 &10.211 &10.572 &11.157 &11.695 &11.757   \\
  22.0 & 9.311 & 9.730 &10.458 &11.032 &11.105 &&  22.0 & 9.810 &10.239 &10.987 &11.567 &11.638   \\
  24.0 & 8.939 & 9.415 &10.287 &10.899 &10.980 &&  24.0 & 9.448 & 9.933 &10.822 &11.442 &11.521   \\
  26.0 & 8.601 & 9.119 &10.117 &10.766 &10.854 &&  26.0 & 9.123 & 9.648 &10.662 &11.317 &11.404   \\
  28.0 & 8.294 & 8.837 & 9.945 &10.627 &10.723 &&  28.0 & 8.832 & 9.380 &10.501 &11.188 &11.283   \\
  30.0 & 8.025 & 8.577 & 9.776 &10.482 &10.585 &&  30.0 & 8.577 & 9.132 &10.341 &11.053 &11.153   \\
  32.0 & 7.797 & 8.354 & 9.620 &10.344 &10.451 &&  32.0 & 8.364 & 8.923 &10.197 &10.925 &11.029   \\
  36.0 & 7.438 & 8.007 & 9.366 &10.110 &10.221 &&  36.0 & 8.038 & 8.609 & 9.973 &10.720 &10.829   \\
  40.0 & 7.171 & 7.756 & 9.174 & 9.931 &10.044 &&  40.0 & 7.800 & 8.386 & 9.808 &10.567 &10.679   \\
  44.0 & 6.952 & 7.551 & 9.013 & 9.778 & 9.893 &&  44.0 & 7.606 & 8.208 & 9.672 &10.439 &10.553   \\
  48.0 & 6.758 & 7.372 & 8.867 & 9.639 & 9.755 &&  48.0 & 7.443 & 8.058 & 9.555 &10.329 &10.445   \\
  52.0 & 6.583 & 7.209 & 8.731 & 9.509 & 9.626 &&  52.0 & 7.299 & 7.925 & 9.450 &10.229 &10.345   \\
  56.0 & 6.419 & 7.055 & 8.601 & 9.383 & 9.501 &&  56.0 & 7.169 & 7.807 & 9.354 &10.138 &10.255   \\
  60.0 & 6.261 & 6.907 & 8.472 & 9.259 & 9.377 &&  60.0 & 7.049 & 7.696 & 9.263 &10.051 &10.169   \\
  64.0 & 6.105 & 6.760 & 8.343 & 9.133 & 9.252 &&  64.0 & 6.936 & 7.592 & 9.176 & 9.968 &10.087   \\
  68.0 & 5.951 & 6.614 & 8.212 & 9.005 & 9.125 &&  68.0 & 6.830 & 7.493 & 9.093 & 9.888 &10.007   \\
  72.0 & 5.796 & 6.466 & 8.078 & 8.874 & 8.995 &&  72.0 & 6.726 & 7.398 & 9.011 & 9.809 & 9.929   \\
  76.0 & 5.636 & 6.313 & 7.938 & 8.737 & 8.858 &&  76.0 & 6.628 & 7.306 & 8.931 & 9.732 & 9.853   \\
  80.0 & 5.470 & 6.153 & 7.789 & 8.592 & 8.713 &&  80.0 & 6.533 & 7.217 & 8.854 & 9.657 & 9.778   \\
  84.0 & 5.300 & 5.989 & 7.637 & 8.441 & 8.563 &&  84.0 & 6.437 & 7.127 & 8.774 & 9.580 & 9.702   \\
   \hline
 \end{tabular} 
 \end{minipage}
 \end{table*}
 
 \begin{table*}
 \begin{minipage}{160mm}
 \contcaption{}
 \centering
 \renewcommand{\footnoterule}{\vspace*{-15pt}}
 \begin{tabular}{crrrrrccrrrrr}
 \hline
 \hline
    $T_{\rm eff}$ & $F_{\rm UV}$ &  $N_{\rm UV}$  &   $V$  &  $J$  &  $H$   &&  $T_{\rm eff}$ & $F_{\rm UV}$ &  $N_{\rm UV}$  &   $V$  &  $J$  &  $H$\\
    ($10^3$ K)    &(mag) &(mag) &(mag) &(mag) &(mag) & & ($10^3$ K) &(mag) &(mag) &(mag) &(mag) &(mag) \\           
\cline{1-6}  \cline{8-13}
    \multicolumn{6}{c}{$1.0\,M_\odot$}  & & \multicolumn{6}{c}{$1.2\,M_\odot$} \\
\cline{1-6}  \cline{8-13}
  10.0 &17.215 &14.156 &13.228 &13.134 &13.069 &&  10.0 &18.175 &14.949 &14.055 &13.944 &13.872   \\
  11.0 &15.660 &13.603 &12.896 &12.968 &12.937 &&  11.0 &16.757 &14.403 &13.726 &13.780 &13.741   \\
  12.0 &14.297 &13.129 &12.638 &12.837 &12.829 &&  12.0 &15.337 &13.931 &13.460 &13.643 &13.627   \\
  13.0 &13.346 &12.745 &12.457 &12.744 &12.751 &&  13.0 &14.341 &13.530 &13.261 &13.539 &13.538   \\
  14.0 &12.587 &12.460 &12.344 &12.683 &12.698 &&  14.0 &13.568 &13.213 &13.123 &13.464 &13.473   \\
  15.0 &12.106 &12.171 &12.219 &12.612 &12.636 &&  15.0 &12.967 &12.958 &13.026 &13.412 &13.429   \\
  16.0 &11.749 &11.911 &12.097 &12.533 &12.567 &&  16.0 &12.560 &12.700 &12.917 &13.349 &13.375   \\
  18.0 &11.185 &11.468 &11.885 &12.382 &12.430 &&  18.0 &11.970 &12.255 &12.705 &13.210 &13.252   \\
  20.0 &10.724 &11.094 &11.701 &12.246 &12.305 &&  20.0 &11.504 &11.883 &12.520 &13.075 &13.129   \\
  22.0 &10.326 &10.765 &11.532 &12.120 &12.187 &&  22.0 &11.111 &11.560 &12.356 &12.953 &13.017   \\
  24.0 & 9.972 &10.465 &11.372 &11.999 &12.075 &&  24.0 &10.766 &11.269 &12.202 &12.837 &12.910   \\
  26.0 & 9.658 &10.189 &11.219 &11.881 &11.965 &&  26.0 &10.460 &10.998 &12.051 &12.721 &12.803   \\
  28.0 & 9.377 & 9.928 &11.063 &11.756 &11.848 &&  28.0 &10.193 &10.748 &11.902 &12.601 &12.691   \\
  30.0 & 9.135 & 9.693 &10.911 &11.627 &11.726 &&  30.0 & 9.964 &10.524 &11.758 &12.478 &12.574   \\
  32.0 & 8.932 & 9.493 &10.775 &11.506 &11.609 &&  32.0 & 9.774 &10.338 &11.630 &12.364 &12.465   \\
  36.0 & 8.624 & 9.197 &10.565 &11.315 &11.423 &&  36.0 & 9.488 &10.063 &11.438 &12.190 &12.296   \\
  40.0 & 8.397 & 8.986 &10.410 &11.172 &11.283 &&  40.0 & 9.276 & 9.866 &11.296 &12.059 &12.169   \\
  44.0 & 8.220 & 8.822 &10.289 &11.058 &11.172 &&  44.0 & 9.110 & 9.715 &11.185 &11.957 &12.069   \\
  48.0 & 8.068 & 8.684 &10.183 &10.959 &11.074 &&  48.0 & 8.971 & 9.588 &11.091 &11.869 &11.983   \\
  52.0 & 7.934 & 8.562 &10.088 &10.870 &10.985 &&  52.0 & 8.849 & 9.478 &11.007 &11.791 &11.907   \\
  56.0 & 7.819 & 8.457 &10.006 &10.792 &10.909 &&  56.0 & 8.741 & 9.380 &10.932 &11.720 &11.837   \\
  60.0 & 7.710 & 8.358 & 9.926 &10.716 &10.834 &&  60.0 & 8.642 & 9.291 &10.862 &11.654 &11.772   \\
  64.0 & 7.611 & 8.267 & 9.853 &10.646 &10.765 &&  64.0 & 8.555 & 9.212 &10.800 &11.596 &11.714   \\
  68.0 & 7.519 & 8.183 & 9.784 &10.581 &10.700 &&  68.0 & 8.475 & 9.140 &10.743 &11.542 &11.662   \\
  72.0 & 7.432 & 8.104 & 9.718 &10.518 &10.637 &&  72.0 & 8.402 & 9.074 &10.690 &11.492 &11.613   \\
  76.0 & 7.349 & 8.028 & 9.654 &10.457 &10.577 &&  76.0 & 8.333 & 9.012 &10.640 &11.445 &11.566   \\
  80.0 & 7.270 & 7.955 & 9.592 &10.398 &10.519 &&  80.0 & 8.268 & 8.953 &10.593 &11.401 &11.522   \\
  84.0 & 7.193 & 7.883 & 9.531 &10.339 &10.460 &&  84.0 & 8.207 & 8.897 &10.547 &11.357 &11.479   \\
  \hline
\end{tabular} 
\end{minipage}
\end{table*}

\section{A sample of bright white dwarfs in the {\it GALEX} all-sky survey}

Table~\ref{tbl-5} lists colours and effective temperatures of known non-DA white dwarfs (DAB, DAO, DB, DO) recovered
in our {\it GALEX}-GSC selection. The reference for the tabulated effective temperature 
and the usual name are also
listed. Table~\ref{tbl-6} provides the same information as Table~\ref{tbl-5} but for
the DA white dwarfs.
 
\begin{table*}
\begin{minipage}{170mm}
\caption{Colours and temperatures of known non-DA white dwarfs in the {\it GALEX} selection.}
\label{tbl-5}
\centering
\renewcommand{\footnoterule}{\vspace*{-15pt}}
\begin{tabular}{ccrrrlcc}
\hline
\hline
 R.A.(J2000)    &  Dec.(J2000) &  $N_{\rm UV}-J$  & $J-H$ &  $T_{\rm eff}$ & Name & Sp. Type & Reference \footnote{
{\sc References}: (1) \citet{wer2008}; (2) \citet{dre1996};  (3) \citet{jor1993}; (4) \citet{heb1996}; 
(5) \citet{koe1994}; (6) \citet{the1991}; (7) \citet{ber1994}.}\\
                &              &     (mag)        & (mag) &   ($10^3$ K)       &      \\
\hline
 00  08  18.2 &$+$51  23  15.1  &$-$1.06 &  $-$0.19& 200.0  & KPD~0005+5106 &  DO  & (1)  \\
 00  41  35.4 &$+$20  09  17.4  &$-$2.42 &  $-$0.09& 115.0  & PG~0038+199   &  DO  & (2)  \\
 01  13  46.7 &$+$00  28  30.9  &$-$2.24 &     0.17&  65.0  & HS~0111+0012  &  DO  & (2)  \\
 02  12  04.8 &$+$08  46  49.8  &$-$2.13 &  $-$0.21&  36.0  & HS~0209+0832  &  DAB & (3)  \\
 05  08  31.0 &$+$01  16  41.3  &$-$2.68 &  $-$0.18&  80:   & HS~0505+0112  &  DAO & (4)  \\
 12  13  56.3 &$+$32  56  33.1  &$-$2.18 &  $-$0.21&  53.0  & HZ~21         &  DO  & (2)  \\
 13  04  32.0 &$+$59  27  33.8  &$-$1.27 &  $-$0.08&  28.8  & GD~323        &  DAB & (5)  \\
 16  47  18.5 &$+$32  28  31.8  &$-$1.45 &  $-$0.08&  25.0  & GD~358        &  DB  & (6)  \\
 23  45  02.9 &$+$80  56  59.0  &$-$1.92 &  $-$0.30&  65.3  & GD~561        & DAO  & (7)  \\
\hline
\end{tabular} 
\end{minipage}
\end{table*}

\begin{table*}
\begin{minipage}{170mm}
\caption{Colours and temperatures of known DA white dwarfs in the {\it GALEX} selection.}
\label{tbl-6}
\centering
\renewcommand{\footnoterule}{\vspace*{-15pt}}
\begin{tabular}{ccrrrlcc}
\hline
\hline
 R.A.(J2000)    &  Dec.(J2000) &  $N_{\rm UV}-J$  & $J-H$ &  $T_{\rm eff}$ & Name & Sp. Type & Reference \footnote{
{\sc References}: (1) \citet{ven1997}; (2) \citet{ven1999}; (3) \citet{ber1992}; (4) \citet{lie2005}; 
(5) \citet{jor1998}; (6) \citet{hil2000}; (7) \citet{dup1998}; (8) \citet{bar1994};
(9) \citet{kaw2007}; (10) \citet{ven1997b}; (11) \citet{sch1995}.}\\
                &              &     (mag)        & (mag) &   ($10^3$ K)       &      \\
\hline
 00  07  32.1 &$+$33  17  27.2  &$-$1.97 &  $-$0.14&  50.8  & GD~2          &  DA  & (1)  \\
 00  39  52.2 &$+$31  32  26.4  &$-$2.05 &  $-$0.12&  52.0  & GD~8          &  DA  & (2)  \\
 00  53  40.5 &$+$36  01  16.6  &$-$1.74 &  $-$0.12&  27.4  & EUVE~0053+360 &  DA  & (1)  \\
 01  04  41.2 &$+$09  49  42.7  &$-$1.40 &  $-$0.05&  25.6  & PG~0102+096   &  DA  & (2)  \\
 01  41  28.7 &$+$83  34  56.9  &$-$1.01 &  $-$0.23&  18.7  & GD~419        &  DA  & (3)  \\
 02  08  03.4 &$+$13  36  24.6  &   0.94 &     0.60&  57.4  & PG~0205+134   &  DA  & (4)  \\
 02  18  48.0 &$+$14  36  06.2  &$-$1.49 &  $-$0.02&  27.2  & PG~0216+144   &  DA  & (1)  \\
 03  04  37.2 &$+$02  56  58.0  &$-$1.68 &  $-$0.40&  35.6  & GD~41         &  DA  & (1)  \\
 03  11  49.1 &$+$19  00  56.1  &$-$0.72 &  $-$0.11&  17.9  & PG~0308+188   &  DA  & (4)  \\
 04  12  43.5 &$+$11  51  47.7  &$-$0.51 &  $-$0.20&  20.8  & HZ~2          &  DA  & (3)  \\
 04  28  39.4 &$+$16  58  12.4  &$-$1.51 &  $-$0.02&  24.4  & EGGR~37       &  DA  & (2)  \\
 05  05  30.7 &$+$52  49  50.1  &$-$1.91 &  $-$0.13&  61.2  & G191 B2B      &  DA  & (1)  \\
 05  10  14.0 &$+$04  38  37.3  &$-$0.99 &  $-$0.20&  20.0  & HS~0507+0434A &  DA  & (5)  \\
 06  23  12.8 &$-$37  41  25.0  &$-$1.51 &  $-$0.11&  61.7  & EUVE~0623$-$376 & DA & (1)  \\
 08  27  05.1 &$+$28  44  02.2  &   1.32 &     0.62&  50.7  & PG~0824+289   &  DA  & (4)  \\
 08  42  53.0 &$+$23  00  25.9  &$-$1.45 &  $-$0.39&  25.0  & PG~0839+232   &  DA  & (4)  \\
 09  42  50.7 &$+$26  00  58.6  &$-$2.36 &  $-$0.21&  67.9  & PG~0824+289   &  DA  & (4)  \\
 09  46  39.1 &$+$43  54  54.9  &$-$0.05 &  $-$0.06&  12.8  & PG~0943+441   &  DA  & (4)  \\
 09  48  46.7 &$+$24  21  25.4  &$-$0.84 &  $-$0.04&  14.5  & PG~0945+246   &  DA  & (4)  \\
 10  36  25.2 &$+$46  08  27.9  &   0.71 &     0.53&  30.2  & GD~123        &  DA  & (1)  \\
 10  44  45.7 &$+$57  44  35.7  &$-$2.01 &  $-$0.19&  30.8  & PG~1041+580   &  DA  & (1)  \\
 10  54  43.4 &$+$27  06  58.4  &$-$1.32 &  $-$0.05&  23.1  & PG~1051+274   &  DA  & (4)  \\
 11  00  34.4 &$+$71  38  03.5  &$-$2.18 &  $-$0.11&  41.2  & PG~1057+719   &  DA  & (1)  \\
 11  07  42.6 &$+$59  58  28.3  &$-$0.87 &  $-$0.05&  17.9  & EGGR~75       &  DA  & (3)  \\
 11  17  03.7 &$+$18  25  58.1  &$-$0.65 &     0.51&  50:   & PG~1114+187   &DA+dMe& (6) \\
 11  26  19.1 &$+$18  39  14.8  &   0.47 &     0.54&  56.9  & PG~1123+189   &  DA  & (1)  \\
 11  32  27.4 &$+$15  17  29.4  &$-$0.68 &  $-$0.05&  16.9  & PG~1129+156   &  DA  & (4)  \\
 11  37  05.2 &$+$29  47  57.6  &$-$1.26 &  $-$0.11&  21.3  & PG~1134+301   &  DA  & (4)  \\
 11  43  59.5 &$+$07  29  04.4  &$-$2.48 &  $-$0.03&  61.8  & PG~1141+078   &  DA  & (4)  \\
 11  45  56.7 &$+$31  49  29.5  &$-$0.52 &     0.03&  14.9  & PG~1143+321   &  DA  & (4)  \\
 11  48  03.2 &$+$18  30  46.3  &$-$1.58 &  $-$0.06&  26.6  & PG~1145+188   &  DA  & (1)  \\
 12  12  33.9 &$+$13  46  26.6  &$-$1.73 &  $-$0.12&  31.9  & PG~1210+141   &  DA  & (4)  \\
 12  36  44.8 &$+$47  55  20.6  &$-$2.21 &     0.02&  56.1  & PG~1234+481   &  DA  & (1)  \\
 13  16  21.8 &$+$29  05  57.8  &   0.81 &     0.57&  50.0  & HZ~43         & DA   & (7) \\
 13  46  01.9 &$+$57  00  33.7  &$-$0.18 &  $-$0.12&  13.4  & PG~1344+573   &  DA  & (4)  \\
 14  10  27.1 &$+$32  08  33.8  &$-$0.89 &  $-$0.09&  18.1  & PG~1408+324   &  DA  & (4)  \\
 16  02  42.1 &$+$57  58  16.0  &$-$0.53 &  $-$0.17&  14.7  & PG~1601+581   &  DA  & (4)  \\
 16  05  21.1 &$+$43  04  36.0  &$-$2.16 &     0.05&  37.0  & PG~1603+432   &  DA  & (4)  \\
 16  38  26.4 &$+$35  00  12.3  &$-$1.80 &  $-$0.01&  37.2  & EUVE~1638+349 &  DA  & (1)  \\
 16  59  48.3 &$+$44  01  04.1  &$-$1.92 &  $-$0.08&  30.5  & PG~1658+441   &  DA  & (4)  \\
 17  13  05.8 &$+$69  31  23.2  &   0.06 &  $-$0.10&  15.6  & EGGR~370      &  DA  & (3)  \\
 17  38  02.8 &$+$66  53  46.7  &$-$2.28 &  $-$0.30&  88.0  & RE~1738+665   &  DA  & (8) \\
 18  00  09.7 &$+$68  35  52.7  &$-$2.23 &  $-$0.43&  46.0  & KUV~18004+6836&  DA  & (1)  \\
 18  47  56.6 &$-$22  19  40.6  &$-$0.47 &     0.62&  31.6  & EUVE~1847$-$223&  DA  & (1)  \\
 19  43  43.8 &$+$50  04  38.9  &$-$1.64 &     0.01&  34.4  & EUVE~1943+500 &  DA  & (1)  \\
 20  10  56.7 &$-$30  13  09.9  &   0.18 &  $-$0.06&  14.8  & LTT~7987      &  DA  & (9) \\
 21  16  53.1 &$+$73  50  41.0  &$-$2.21 &  $-$0.41&  54.7  & KUV~21168+7338&  DA  & (1)  \\
 21  24  58.1 &$+$28  26  03.2  &$-$1.25 &  $-$0.21&  53.0  & EUVE~2124+284 &  DA  & (10) \\
 21  52  25.2 &$+$02  23  18.6  &$-$0.97 &  $-$0.08&  18.2  & EGGR~150      &  DA  & (3)  \\
 22  58  48.2 &$+$25  15  43.4  &   1.29 &     0.49&  22.2  & GD~245        &DA+dMe& (11)\\
\hline
\end{tabular} 
\end{minipage}
\end{table*}

\label{lastpage}
\end{document}